\documentclass[aps,prl,longbibliography,twocolumn,10pt,superscriptaddress,amsmath,amssymb]{revtex4}
\usepackage{amsmath}
\usepackage{amsfonts}
\usepackage{amsbsy}
\usepackage{amssymb}
\usepackage{graphicx}
\usepackage{textcomp}
\usepackage{subfigure}
\usepackage{xcolor}
\usepackage{mathrsfs}
\usepackage{bm}
\usepackage{ulem}
\usepackage{float}
\usepackage[colorlinks,linkcolor=red,citecolor=blue,filecolor=magenta,
urlcolor=cyan]{hyperref}
\newcommand{\nn}{\nonumber}

 \newcommand{\bs}{\boldsymbol}

\allowdisplaybreaks
  
\begin{document}
\title{Dynamical torques from Shiba states in $s$-wave superconductors}

\author{Archana Mishra}
\email{mishra@MagTop.ifpan.edu.pl}
\affiliation{International Research Centre MagTop, Institute of Physics, Polish Academy of Sciences,
Aleja Lotnikow 32/46, PL-02668 Warsaw, Poland}
\author{So Takei}
\affiliation{The Graduate Center, City University of New York, 365 Fifth Avenue New York, NY 10016 USA}
\author{Pascal Simon}
\email{pascal.simon@u-psud.fr}
\affiliation{Universit\'e Paris-Saclay, CNRS, Laboratoire de Physiques des Solides, 91405, Orsay, France}
\author{Mircea Trif}
\email{mtrif@MagTop.ifpan.edu.pl}
\affiliation{International Research Centre MagTop, Institute of Physics, Polish Academy of Sciences,
Aleja Lotnikow 32/46, PL-02668 Warsaw, Poland}

\date{\today}
\begin{abstract}
Magnetic impurities inserted in a {\it s}-wave superconductor give rise to spin-polarized in-gap states called Shiba states. We study the back-action of these induced states on the dynamics of the classical moments. We show that the Shiba state pertains to both reactive and dissipative torques acting on the precessing classical spin that can be detected through ferromagnetic resonance  measurements. Moreover, we highlight the influence of the bulk states as well as the effect of the finite linewidth of the Shiba state on the magnetization dynamics. Finally, we demonstrate that the torques are a direct measure of the even and odd frequency triplet pairings  generated by the dynamics of the magnetic impurity. Our approach offers non-invasive alternative to the STM techniques used to probe the Shiba states.  
\end{abstract}
\maketitle
{\it Introduction} $-$ The quest for Majorana fermions is driven by their promise as a building block for a fault tolerant topological quantum computer \cite{sarma2015}. Magnetic impurities in superconductors have been a prime area of research for realizing topological superconductors that can host such exotic quasi-particles \cite{choy2011,nakosai2013,nadj2013,braunecker2013,klinovaja2013,vazifeh2013,pientka2013,pientka2014,poyhonen2014,heimes2014,reis2014,weststrom2015,peng2015,rontynen2015,braunecker2015,zhang2016,hoffman2016,neupert2016,kimme2016,kaladzhyan2017,andolina2017}. 
The elementary unit for such a chain is a single magnetic impurity inserted in a {\it s}-wave superconductor, which  can bind spin polarized sub-gap energy electrons in the so called Shiba states \cite{shiba1968}. 

The Shiba impurity states have been well studied theoretically \cite{yu1965, shiba1968,rusinov1969} and experimentally revealed by scanning tunneling microscopy (STM) technique. Unfortunately, such systems are hard to tune once in the superconductor, which drastically reduces the ability to explore various topological regimes. Driving the impurities,  however, can result in the ability to achieve such a feat dynamically where the precessing frequency acts as the knob to control the topological transition \cite{kaladzhyan2017}. Spin pumping and spin transfer torques in ferromagnets are just a few phenomena that pertain to magnetization dynamics  \cite{stiles2006, sankey2006} and which are staple dynamical methods for manipulating, transporting, and detecting spins in magnetic systems. Recently, it has been shown theoretically that such dynamics results in controllable  shifts in the Shiba energies that show up  as features in the differential conductance  in transport measurements \cite{kaladzhyan2017a}. Thus, dynamical magnetic impurities  are  promising  platform  for engineering topological superconductor \cite{kaladzhyan2017}.  
\begin{figure}
    \includegraphics[width=0.8\linewidth,trim=95mm 0mm 0mm 0mm,clip]{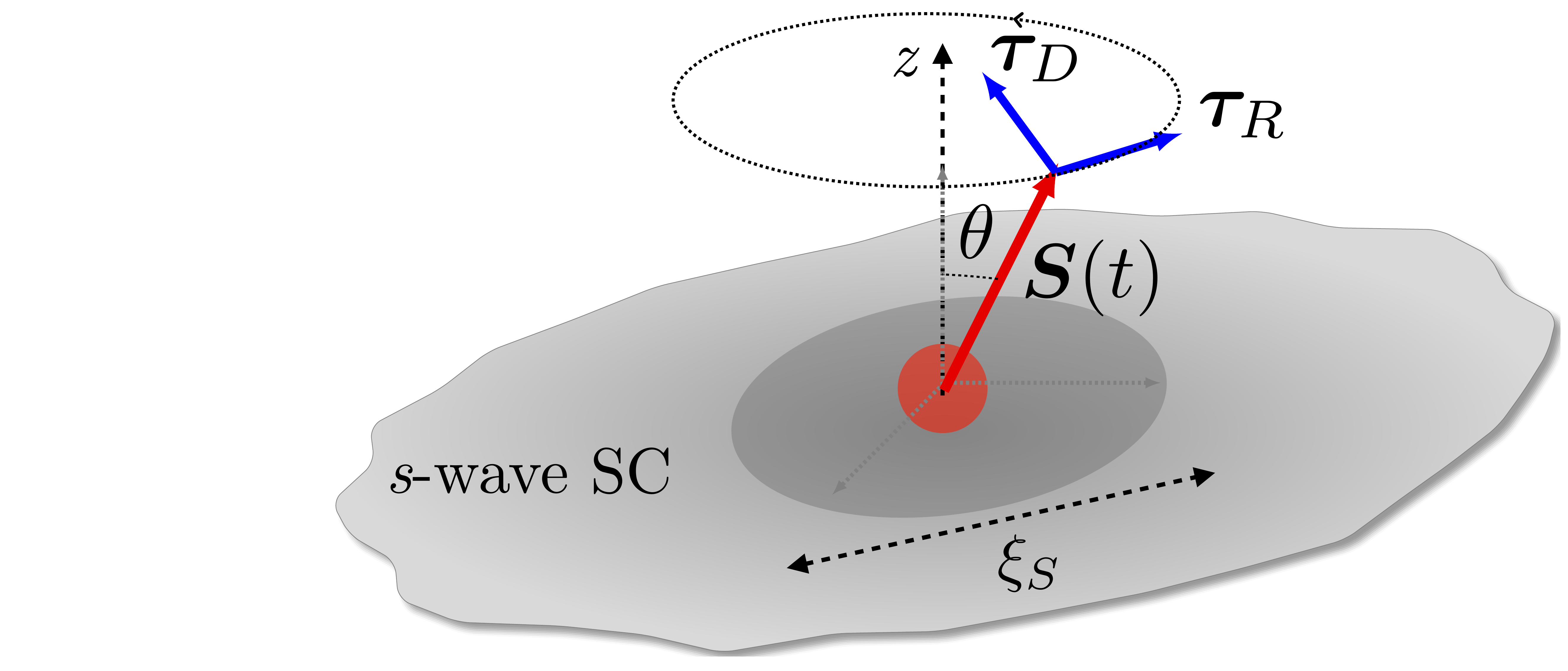}
    \caption{Sketch of the combined system. A classical precessing spin (red) at angle $\theta$ with respect to the $z$-axis is coupled via exchange interaction to an $s$-wave superconductor (grey). A localized Shiba state of size $\xi_S$ is formed underneath which is spin polarized and can act back on the classical spin precession. Both reactive (${\bs \tau}_R$) and dissipative (${\bs \tau}_D$) torques are present which affect the dynamics of the classical spin.}
    \label{fig:1}
\end{figure}
In this paper, we take a step forward and investigate a single  time-dependent magnetic impurity of size $S$ in an {\it s}-wave superconductor (SC), in particular the  back-action  effects  of  the ``stirred" electrons in the SC  on the  spin dynamics. In the adiabatic limit, we find a universal reactive torque pertaining to the  Shiba state that is  geometrical in nature:
\begin{equation}
{\bs \tau}_R(t)=(n_S-1/2)\,F_{s}[\bs{n}(t)]\,\dot{\bs n}(t)\,,
\label{eq:1}
\end{equation}
where $n_S$ is the occupation of the Shiba state, $F_s[\bs{n}(t)]=S\bs{B}\cdot\bs{n(t)}$  is the radial Berry curvature of the Shiba state, $\bs{B}$ is the magnetic field, and ${\bs n}(t)={\bs S}(t)/S$ is the precessing classical spin direction. Changing $n_S$  is equivalent to effectively changing the classical spin length as $S\rightarrow S-(n_S-1/2)$. Berry-phase induced torques and their effects on classical spins in normal metals have been investigated previously in several important works  \cite{millis1995,stahl2017,elbracht2020,suresh2020,bajpai2020}. However, the origin of the torque in all those instances is different from that described here pertaining to precessing spins in superconductors. In Fig.~\ref{fig:1} we show a sketch of the combined dynamical spin-SC system, and highlight the reactive (${\bs \tau}_R$) and dissipative (${\bs \tau}_D$) torques that act back on the classical spin, respectively.  The rest of the paper is dedicated to derive  Eq.~\ref{eq:1} microscopically,  to take into account a  finite Shiba line-width,  which in turn leads to a  dissipative  torque (${\bs \tau}_D$), as well as for the effects of the bulk (non-localized states in the continuum) states in the superconductor. Further, we show how these torques are a measure of various even- and odd-frequency triplet SC pairings induced by the dynamics. 

{\it Model Hamiltonian} $-$ The model Hamiltonian for the dynamical system in Fig.~\ref{fig:1}, which describes both 2D and 3D setups, can be written as $H_{\rm tot}(t)=(1/2)\int d{\bs r}\Psi^\dagger({\bs r}) H_{\rm BdG}(t)\Psi({\bs r})$ where the Bogolioubov de Gennes Hamiltonian 
reads:
 \begin{align}
 \label{eq c:2}
H_{\rm BdG}(t)=&H_0+V_i(t)\,,\\
H_0=&\epsilon_p\tau_z+\Delta\tau_x\,,\nn\\
V_{i}(t)=&-J{\bs S}(t)\cdot{\bs\sigma}\,\delta({\bs r})\,,\nn
\end{align}
with  $H_0$ and $V_{i}(t)$ being the bare Bogolioubov de Gennes Hamiltonian for the {\it s}-wave superconductor and its coupling to the classical magnetic impurity respectively written in the Nambu basis $\Psi({\bs r})=[c_{\uparrow}({\bs r}), c_{\downarrow}({\bs r}),  c^\dagger_{\downarrow}({\bs r}), -c^\dagger_{\uparrow}({\bs r})]^T$. Also, ${\bs\sigma}=(\sigma_x,\sigma_y,\sigma_z)$ and ${\bs\tau}=(\tau_x,\tau_y,\tau_z)$ are the Pauli spin matrices in the spin and particle hole subspace respectively. The spectrum of free electrons is given as  $\epsilon_p=p^2/2m-\mu$, where $m$, $p$, and $\mu$ are the electron mass, momentum,  and chemical potential, respectively.  $\Delta$ is the superconducting order parameter and $J$ defines the coupling between the classical spin ${\bs S(t)}=S(\sin\theta\cos\phi,\sin\theta\sin\phi,\cos\theta)$ and the electrons in the superconductor. In the following,  we assume circular precession, i.e. $\phi=\Omega t$, with $\Omega$ and  $\theta$ representing the precession frequency and angle the classical spin makes with $z$-axis, respectively. The magnetic impurity generates also a scalar potential but, for simplicity,  we neglect such contribution in this work. The dynamics of the classical spin is described by the Landau–Lifshitz-Gilbert (LLG) equation: 
\begin{align}
\label{eq:2}
\dot{\bs S}(t)&=-{\bs S}(t)\times\left(\gamma{\bs B}(t)-\langle{\bs\sigma}(t)\rangle+\beta\dot{\bs S}(t)\right)\,,
\end{align}
where $\gamma$ and $\beta$ are the gyromagnetic coupling and the Gilbert damping respectively, and ${\mathcal{\bs\tau}}_{\rm R}(t)=J{\bs S}(t)\times\langle{\bs\sigma}(t)\rangle$  is the total torque acting on the classical spin by the superconductor, with $\langle{\bs \sigma}(t)\rangle$ being the spin expectation value in the superconductor at the position of the impurity in the steady state.  This term  can change both the resonance frequency and the Gilbert damping.  Note that in typical setups ${\bs B}(t)=B_0{\bs z}+{\bs B}_\perp(t)$, with $|{\bs B}_\perp(t)|\ll B_0$ (small angle precession) and $\gamma B_0\equiv \Omega_0$ defines the resonance frequency in the absence of the Shiba states. 

{\it Rotating-frame description} $-$ It is convenient to  analyze the dynamics by using the rotating wave description approach.  
Due to the circular precession of the magnetic impurity, the symmetry of the system allows us to perform a unitary transformation $U(t)$ that renders the problem fully static. Hence, we can write $\Psi({\bs r},t)=U(t)\Phi({\bs r})e^{-iEt}$ such that the time independent Schrodinger equation can be written as 
$H_{\rm rot}\Phi({\bs r})=E\Phi({\bs r})\,$
with $H_{\rm rot}=U^\dagger(t)H_{\rm tot}(t)U(t)+i\dot{U}^\dagger(t)U(t)$ or
\begin{equation}
H_{\rm rot}=H_{\rm tot}(0)-b\sigma_z\,,
\label{eq:3}
\end{equation}
where $U(t)=\exp{(-ib\sigma_zt)}$ and $b=\Omega/2$ is the fictitious magnetic field perpendicular to the plane of the superconductor.

In the absence of precession, a magnetic impurity in a {\it s}-wave superconductor gives rise to Shiba state
within the superconducting gap at energy, 
$E_S=\Delta(1-\alpha^2)/(1+\alpha^2)$ and $\alpha=\pi\nu_0JS$  is the dimensionless impurity strength in terms of the normal phase density of states $\nu_0$. For finite precession of the impurity, the coherence peaks split due to the fictitious magnetic field, therefore it becomes easier to break the Cooper pair and thus lower the energy of the excitation.

When the impurity spin precesses, a general  solution to the eigenvalue problem has a complicated form [see supplementary material (SM) \cite{SM} for details ], but in the deep Shiba limit,  $\alpha\approx 1$ and in the adiabatic regime, $b/\Delta\ll1$, the effective Shiba energy acquires the simple expression $E_S'\approx E_S-b\cos\theta$ in  leading  order in $b/\Delta$. The corresponding wave-function is $|\Phi_S\rangle\approx|\Phi^0_{S}\rangle+(X/2)\sin\theta|\Phi^1_{S}\rangle$, where $|\Phi^0_{S}\rangle=\left[\cos{(\theta/2)},\sin{(\theta/2}),\cos{(\theta/2)},\sin{(\theta/2})\right]^T$ and $|\Phi^1_{S}\rangle=\left[\sin{(\theta/2)},-\cos{(\theta/2}),\sin{(\theta/2)},-\cos{(\theta/2})\right]^T$ scaled up to a normalization factor $1/\sqrt{N}$ where $N=(1+\alpha^2)^2/(2\pi\nu_0\alpha\Delta)$ \cite{SM}, and $X=\frac{b}{\Delta}\frac{(1+\alpha^2)^2}{4\alpha^2}\ll1$.
The dynamics induces a coupling between the static Shiba state to its spin partner in the continuum. 
Considering either the electron or hole component, the result can be interpreted in the context of an adiabatically driven spin $1/2$ particle in an effective magnetic field $B_{\rm eff}=b/X$.
The average spin at the site of the impurity pertaining to the Shiba state can be calculated from the above renormalized wavefunctions $|\Phi_S\rangle$ in leading order in $b/\Delta$, and accounting for their occupation \cite{SM}:
\begin{equation}
    \langle{\bs\sigma}_S(t)\rangle\approx\frac{2n_S-1}{N}\left[(1-X\cos\theta){\bs n}(t)+X{\bs z}\right]\,,
    \label{eq:5a}
\end{equation}
where $n_S\equiv n_S(b,\theta)$ is the occupation number for the Shiba state which itself can depend on the driving.
The misalignment of the Shiba state spin and the classical moment is due to the competition between the local exchange field and the fictitious global magnetic field acting along the ${\bs z}$ direction.
Using Eq.~\eqref{eq:5a}, the universal torque can be evaluated as presented in Eq. \ref{eq:1}, with $F_s[{\bs n}(t)]=1/2$ corresponding to the Berry curvature of the effective spin $1/2$. However, this contribution is due to the purely isolated Shiba state in the absence of any relaxation channels.  To account for  the full out-of-equilibrium properties, including the bulk states and account for the various dissipation effects, in the following we analyze the dynamical problem  by employing the Green's function (GF) technique.

{\it GF approach} $-$ The bare retarded GF of the superconductor in the rotating frame is 
\begin{equation}
\label{eq:5}
\tilde G_0(\omega)=-\frac{\pi\nu_0}{2}\sum_{\sigma=\pm1}\frac{\omega+\sigma b+\Delta\tau_x}{\sqrt{\Delta^2-(\omega+\sigma b)^2}}(1+\sigma\sigma_z)\,,
\end{equation}
where $\omega=\omega+i0^+$ and $\sigma=+1(-1)$ for $\uparrow(\downarrow)$. The coupling to the impurity spin in the rotating frame can be accounted for via the Dyson's equation that relates the full GF to the bare one, or  $[\tilde G^{R}(\omega)]^{-1}=\tilde G^{-1}_0(\omega)-V_i(0)+i\Gamma$. Here, $\Gamma$ is a phenomenological Dynes broadening added to the self energy that accounts for the relaxation processes in the superconductor. That in turn allows us to write
\begin{align}
\label{eq:6}
\tilde G^R(\omega)=&\frac{\pi\nu_0}{D(\omega,\alpha,b,\theta)}M(\omega,\alpha,b,\theta)\,,
\end{align} 
where 
 
$M(\omega,\alpha,b,\theta)=M_0+\bs{M}_1\cdot\bs{\sigma}+\tau_x\otimes(M_2+\bs{M}_3\cdot\bs{\sigma})$
is a $4\times4$ matrix and $D(\omega,\alpha,b,\theta)=\omega_1\omega_2(1+\alpha^4-2\alpha^2\cos^2\theta)-2\alpha\cos\theta(1-\alpha^2)[(b+\omega)\omega_2+(b-\omega)\omega_1]$, where $\omega_1=\sqrt{\Delta^2-(\omega+b)^2}$ and $\omega_2=\sqrt{\Delta^2-(\omega-b)^2}$. 
The full Shiba state energy is found from the solutions of 
$D(\omega,\alpha,b,\theta)=0$ \cite{SM}.
In the absence of precession, the lesser GF, $\tilde G^<(\omega)=n_F(\omega)[\tilde G^A(\omega)-\tilde G^R(\omega)]$ where $n_F(\omega)=[\exp(\beta\omega)+1]^{-1}$ is the Fermi distribution function with $\beta=1/k_BT$, $k_B$ being the Boltzmann constant and $T$ is the temperature. The advanced GF instead satisfies 
$\tilde G^A(\omega)=[\tilde G^{R}(\omega)]^\dagger$.  In this work, we are not considering the microscopic mechanisms behind $\Gamma$, but rather focus on its manifestations on the ferromagnetic resonance (FMR) signal. 
For finite precession and in the rotating frame  the fictitious magnetic field $b$ leads to a spin dependent shift in the Fermi distribution function  and is no longer an identity operator: \begin{align}
\label{eq:8}
\tilde{n}_F\equiv\tilde{n}_F(\omega,b)=&f_0(\omega,b)+f_s(\omega,b)\sigma_z\,,
\end{align}
where $f_{0,s}(\omega,b)=\left[n_F(\omega+b)\pm n_F(\omega-b)\right]/2$. Note that $\tilde{n}_F(\omega,b)$ does not commute with $\tilde G^R(\omega)$, and  the lesser GF is found as \cite{teber2010} (also \cite{SM}):
\begin{equation}
\tilde G^<(\omega)=\tilde{n}_F\tilde G_S^A-\tilde G_S^R \tilde{n}_F+\tilde G_S^R(V_i\tilde{n}_F-\tilde{n}_FV_i)\tilde G_S^A\,.
\label{eq:9}
\end{equation}
The instantaneous spin expectation value at the position of the impurity in the rotating frame is 
\begin{equation}
\langle\tilde{\bs\sigma}({\bs 0})\rangle=\frac{-i}{2\pi}\int_{-\infty}^\infty d\omega\,{\rm Tr}\left[\left({\bs\sigma}\otimes\frac{1+\tau_z}{2}\right)\tilde G^<(\omega)\right]\,,
\label{eq:7}
\end{equation}
which contains both the in-gap (Shiba) and the bulk (continuum of states) contributions, respectively.
Here,  the  $(1 +\tau_z)/2$  term  is  introduced  in  order  to  account  for  only  the  electron  components.
For a static impurity spin, the imaginary part of the integrand in Eq.~\eqref{eq:7} is a Lorentzian  function located at the Shiba energies and the expectation value of the spins is non-zero for a finite Shiba linewidth. The contribution to $\langle\tilde{\bm{\sigma}}({\bs 0})\rangle$ is only due to the Shiba states and it points  along the classical spin direction resulting in zero net torque in the absence of precession, as expected. For the dynamic case, the spin expectation value has contribution from both the Shiba and the bulk states,  as discussed below.

\begin{figure}
    \includegraphics[width=\linewidth]{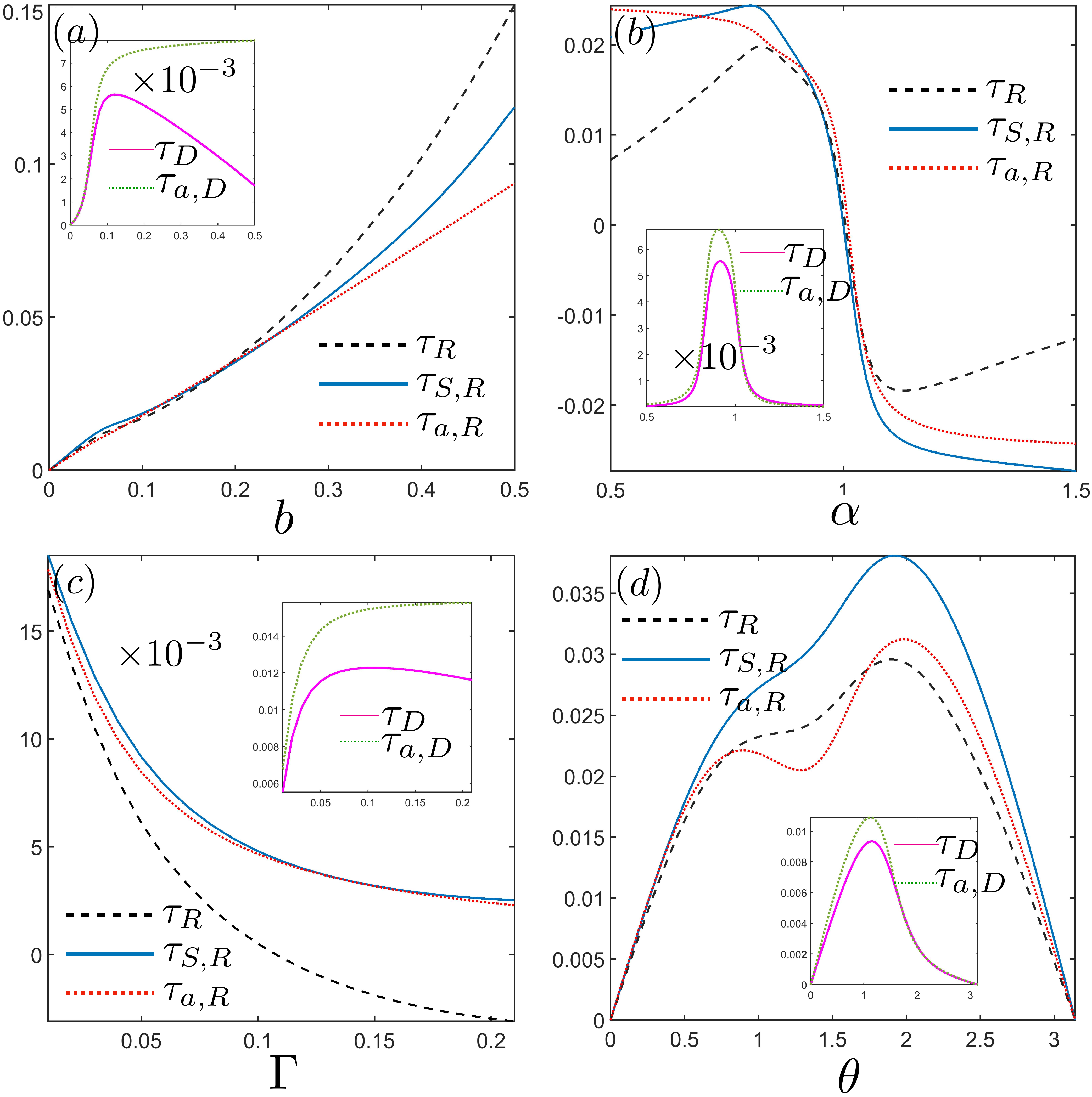}
       \caption{Variation of the reactive and dissipative torques. Main: full Shiba $\tau_{S,R})$ (blue solid line), approximate Shiba  $\tau_{a,R}$ (red dotted line) and  total $\tau_{R}$ (black dashed line) reactive torques, respectively  as a function of  b,  $\alpha$,  $\Gamma$, and  $\theta$ [shown in $(a)$, $(b)$,  $(c)$, and  $(d)$, respectively]. We assumed $\alpha=0.9$ [$(a)$, $(c)$, and $(d)$], $\theta=\pi/6$ [$(a)$, $(b)$, and $(c)$], $\Gamma=0.01$ [$(a)$, $(b)$, and $(d)$], and  $b=0.1$ [$(b)$, $(c)$, and $(d)$]. The insets in the plots show the dependence of the full $\tau_{D}$ (full magenta) and approximate $\tau_{a,D}$ (dotted green) dissipative torques, respectively on the corresponding parameters, with the other values being the same as for the main plots.}
      \label{fig:2}
\end{figure}

{\it Effect of the Shiba states} $-$ The in-gap Shiba contribution stems from the range of integration $\omega\in[-\Delta+b,\Delta-b]$ in Eq.~\ref{eq:7}. While later on we will evaluate this term fully numerically, let us next consider the deep Shiba limit \cite{ruby2015,kaladzhyan2016,kaladzhyan2017} and $b/\Delta\ll1$ so that these states are well separated from the bulk.  
As mentioned before, the average spin of the Shiba state for precessing case is no longer along the classical spin direction ${\bs n}$  \cite{SM} and hence, a finite torque acts on the classical spin due to this deviation which in turn will affect its dynamics. After lengthy but straightforward calculations, we   find compact analytical expressions for the spin expectations values that pertain to the reactive and dissipative torques, respectively, and at $T\rightarrow0$:
\begin{align}
 \label{eq:11}
    \langle\tilde{\sigma}_{a,R}\rangle&\approx  -b(n_++n_-\cos{\theta});\,\,\,
       \langle\tilde{\sigma}_{a,D}\rangle\approx\Gamma_S\,n_-\,,\\
    n_{\pm}&=\frac{1}{2\pi}\sum_{s=\pm1}s^{p}\arctan\left(\frac{E_S'+s b}{\Gamma_S}\right)\,,
  \end{align}
where $\Gamma_S=(2/N)\,\Gamma$ is the effective Shiba linewidth and $p=0(1)$ for $n_{+(-)}$. Eqs.~\ref{eq:11} supplemented with the plots in Fig.~\ref{fig:2} is the main finding in this work. In the rotating frame,  we can write the torque stemming from the Shiba state only as
\begin{align}
{\bs \tau}_S=&\langle\tilde\sigma_{a,R}\rangle{\bs n}\times {\bs z}+\langle\tilde\sigma_{a,D}\rangle{\bs n}\times({\bs z}\times{\bs n})\,,
\label{eq:12}
\end{align}
where the first (second) term correspond to the reactive (dissipative) torque $\bm{\tau}_{S,R}$ ($\bm{\tau}_{S,D}$).
We see that both the reactive and dissipative torques vanish for either $b=0$ or $\theta=0,\pi$, as expected. Moreover, the reactive torque can be casted in the form shown in  Eq.~\ref{eq:5a}, by identifying $n_S\equiv1/2- (n_++n_-\cos{\theta})$ as the occupation number of the Shiba state (see SM).     

 A few comments are in place. In order to extract the dissipative torque, we accounted  for the linewidth $\Gamma$ not only in the denominator (that reflects the Shiba state lifetime through $\Gamma_S$), but also in the numerator $M_0$ and ${\bs M}$.  Our expansion goes beyond the effective Shiba approximations discussed in, for example, Ref.~\cite{ruby2015}, where they neglect contributions of $\Gamma$ in the numerator, and which would naively lead to a vanishing dissipative torque. Our theory shows, we believe, one of the first instances where such an effective approach is not sufficient for the case of Shiba states in superconductors.

{\it Bulk effects.}$-$ The above torques account only for the in-gap contributions stemming  from the Shiba poles, while the full average spin value at the impurity can be evaluated only numerically.  Interestingly, in the general case, the  integrand in Eq.~\ref{eq:7}  is non-zero even when $\Gamma=0$ for $|\omega|>\Delta-b$. Hence, the precession of the impurity can result in a finite contribution of the  bulk states to the dynamical torques,  absent in the static case.
The total spin expectation value can be written as  $\langle\tilde{\bs \sigma}\rangle=\langle\tilde{\bs \sigma}_B\rangle+\langle\tilde{\bs\sigma}_S\rangle$ (and similarly for the torques),  denoting sum of the bulk and Shiba contributions, respectively. 
In Fig.~\ref{fig:2} we show the total reactive torque ($\tau_R$) and the Shiba parts of the reactive torque ($\tau_{S,R}$) as a function of various parameters, such as  $b$, $\alpha$, $\Gamma$  and $\theta$, calculated numerically. The difference between $\tau_R$ and $\tau_{S,R}$ accounts for the bulk contribution to the reactive torque. The plot of the dissipative torque   ($\tau_D$), which originates only from the Shiba state, is showed in the insets in Fig.~\ref{fig:2}. To compare the full numerics  with the analytical results describing the Shiba contribution, we also plot the reactive ($\tau_{a,R}$) and dissipative torques $\tau_{a,D}$ calculated from the
Eq.~\eqref{eq:11} and shown by the dotted line in Fig.~\ref{fig:2}. We see a very good agreement between the analytic expressions and the Shiba contribution obtained from the numerics for small $b$ (Fig.~\ref{fig:2}a) and in the deep Shiba limit (Fig.~\ref{fig:2}b). Furthermore, in this limit,  $\tau_R$ and $\tau_{S,R}$ coincide, proving  that the Shiba state is responsible for the reactive torque, while the bulk contribution tends to zero. Moving away from the deep Shiba limit, there is finite bulk contribution to the reactive torque (see Fig.~\ref{fig:2}b). Fig.~\ref{fig:2}(c) and (d) shows the behavior of the dynamical torques as a function of the Dynes broadening factor $\Gamma$ and $\theta$, respectively. By increasing  $\Gamma$  the difference between $\tau_R$ from $\tau_{S,R}$ increases also. Note that since ${\bs \tau}_D\propto{\bs n}\times\dot{\bs n}$ is always positive, as per LLG equation given in Eq.~\eqref{eq:2} it corresponds  to a damping-like/ dissipation torque. From Fig.~\ref{fig:2}(b), $\tau_D$ peaks at $\alpha$ corresponding to $E_S'=0$ for the given parameters as can be seen from the analytic expression Eq.~\ref{eq:11}. Experimentally, that would result in a strong enhancement of the FMR linewidth. 

{\it Unconventional pairing} $-$  The occurrence of odd frequency superconductivity has been  recently discussed  for a static impurity \cite{kuzmanovski2019,perrin2019}. Here we show that spin precession leads to generation of unconventional pairing that is directly related to the experimentally accessible dynamical torques.
Similar to Ref.~\cite{perrin2019}, we consider the adiabatic deep Shiba limit and expansion of the numerator and denominator of $\tilde G^R_\pm(\omega)$ in zeroth and first order in $\Gamma$, respectively, we can write
 \begin{align}
\tilde G^R_\pm(\omega)\approx&\frac{\pi\nu_0(\tau_0\pm\tau_x)}{\omega\mp E_S'+i\Gamma_S}(M_0\pm{\bs M}\cdot{\bs\sigma})\,,
\label{eq:10a}
\end{align}
where $M_0$ [${\bs M}=(M_x, 0, M_z)$] is a scalar (vector) that depends on the precession frequency $b$, angle $\theta$ and impurity strength 
$\alpha$. Eq.~\ref{eq:10a} is the second main result of our work. In this limit,  the anomalous part of the retarded GF in rotating frame, $\tilde F(\omega)$, corresponds to the term $\propto\tau_x$ in Eq.~\eqref{eq:10a} (see SM). 
 The reactive torque for a given frequency $\omega$ originating from the Shiba state (which when integrated gives the total torque) can then be written in terms of the anomalous pairing as follows: 
\begin{align}
    \!\!\!\tau_{S,R}(\omega)=&\,\mbox{Im}\left[\tilde F_{o}(\omega) f_s(\omega)\sin\theta-\tilde F_{e}(\omega)f^o_0(\omega)\right]\,,
       \label{eq:16}
\end{align}
where the even  and odd frequency  triplet pairing components above  are defined as $F_{o}(\omega)=\sum_\sigma F_{\sigma\bar{\sigma}}^{o}(\omega)$, $F_e(\omega)=\sum_\sigma F_{\sigma\sigma}^{e}(\omega)$ with  $\tilde F_{\sigma\sigma'}^{e/o}(\omega)=\tilde F_{\sigma{\sigma'}}(\omega)\pm \tilde F^*_{\sigma{\sigma'}}(-\omega)$. Morerover, $f^{o}_0(\omega)=f_0(\omega)-1/2$ while $f_s(\omega)$ is even under $\omega\rightarrow-\omega$.
As seen from Eq.~\eqref{eq:16}, the reactive torque is generated by two types of pairing: ($i$) a $b$ dependent induced triplet pairing and ($ii$) an odd frequency pairing term independent of $b$ (in leading  order in $b$). Measuring the reactive torque experimentally through FMR can act as a probe to such unconventional pairing.
Furthermore, when the Shiba state is completely filled or empty ($f_s=0$), a finite reactive torque establishes the presence of precession induced triplet superconducting pairing. The total reactive torque can be obtained by integrating over $\omega$. Such dynamical generation of triplet pairing and its connection to the torques can be utilised to both manipulate and detect the topological phase diagram of a chain of Shiba impurities and eventually of the emergent Majorana fermions by standard spintronic techniques \cite{valenzuelaRMP15}. Nevertheless, such a study is beyond the scope of this paper and it is left for future work.

Finally, let us give some estimates for the possible FMR frequency shift $\delta\Omega_r\sim\Omega_0/(4S+1)$ of a spin $S$ impurity  (details in SM), where $\Omega_0$ is the bare Larmor frequency. For example, considering the experimental system of Moire patterns of adsorbates on a conventional SC (Pb) with SC gap $\Delta\approx3\,\mbox{meV}$ and $S=1$, described in Ref.~\cite{hatter2017}, $\delta\Omega_r=20\,\mbox{GHz}$ for $\Omega_0=100\,\mbox{GHz}<\Delta$. Similarly, for an impurity spin $S=5/2$ corresponding to the transition metals \cite{vzitko2018}, $\delta\Omega_r\approx9\,\mbox{GHz}$ which is within the current experimental resolution \cite{willke2018}.

{\it Conclusions and outlook} $-$ In conclusion, we have investigated the dynamical torques acting on a clasical spin $S$ precessing in an $s$-wave superconductor.
We found that the torques originate  both from the Shiba and the bulk states, with the former contribution having a geometrical (Berry-phase) origin that shifts the FMR frequency.  Using various methods, we showed that a finite linewidth of the Shiba state results in an extra damping of the precession of the classical spin.  Finally, we showed that  classical spin precession generates  unconventional superconducting pairings which is directly reflected into the dynamical torques. Our results offer a non-invasive alternative to the usual STM techniques to address and manipulate the Shiba states, could be relevant for arrays of dynamical magnetic impurities in superconductors.  While manipulating the atomic spin in metal substrate using the STM and electron spin resonance has been experimentally realized \cite{natterer2017,yang2019}, we expect our results will be pertinent and useful to extend the study to magnetic moment in superconductor.

{\it Acknowledgments} $-$ This work was supported by the International Centre for Interfacing Magnetism and Superconductivity with Topological Matter project (AM and MT), carried out within the International Research Agendas program of the Foundation for Polish Science co-financed by the European Union under the European Regional Development Fund. We would like to thank Silas Hoffman, Marco Aprili, and Hervé Aubin for the interesting and fruitful discussions.

\newpage
\onecolumngrid
\section*{Supplementary Material for ``Dynamical torques from Shiba states in $s$-wave superconductors"}
%
%
\maketitle
\subsection{Wave function approach}

\subsubsection{Average spin of Shiba state}

The BCS Hamiltonian in the presence of the precessing impurity reads:
\begin{align}
H_{BCS}(t)=\sum_{\alpha,\beta}\int d{\bs r}\,c^\dagger_\alpha({\bs r})\left[\left(\frac{p^2}{2m}-\mu\right)\delta_{\alpha\beta}-J{\bs S}(t)\cdot({\bs \sigma})_{\alpha\beta}\delta({\bs r})\right]c_\beta({\bs r})+\sum_{\alpha}\int d{\bs r}\left[\Delta\, c^\dagger_\alpha({\bs r})c^\dagger_{\bar{\alpha}}({\bs r})+{\rm h. c.}\right]\,,
\end{align}
where $c^\dagger_\alpha({\bs r})$ ($c_\alpha({\bs r})$) is the creation (annihilation) fermionic operator at position ${\bs r}$ and spin $\alpha$, $\bar{\alpha}=-\alpha$. Here, $S(t)=S(\sin{\theta}\cos{\Omega t}, \sin{\theta}\sin{\Omega t}, \cos{\theta})$. 
Following Ref.~\onlinecite{TauberPRB10}, we can  switch to a  rotating frame using a unitary transformation $\mathcal{U}(t)$ (details described later) leading to a fully static problem. 
The resulting Hamiltonian in rotating frame is:
\begin{align}
\tilde{H}_{BCS}=H_{BCS}(0)-b\sum_{\alpha}\int d{\bs r}\,c^\dagger_\alpha({\bs r})(\sigma_{z})_{\alpha\alpha}c_\alpha({\bs r})\,,
\end{align}
where $b=\Omega/2$. The spin operator associated with the electrons in the superconductors at ${\bs r=0}$ can be written as:
\begin{equation}
\hat{\bs \sigma}({\bs 0})=\sum_{\alpha,\beta}c_\alpha^\dagger({\bf 0})({\bs\sigma})_{\alpha\beta}c_\beta({\bf 0})\,.
\end{equation} 
We can perform  a canonical transformation to  write the fermionic field operators as follows  \cite{BalatskyRMP06}:
\begin{equation}
c_{\alpha}({\bf 0})=\sum_n \left[u_{n\alpha}({\bf 0})\gamma_n-\alpha v^*_{n\alpha}({\bf 0})\gamma^\dagger_n\right]\,,
\label{seq:9}
\end{equation}
where $\gamma_n (\gamma^\dagger_n)$ is the annihilation (creation) operator of the Bogoliubov quasiparticles in state $n$ corresponding to the eigenstates of the effective Hamiltonian in the rotating frame, while $u_{n\alpha}({\bs r})$ $\left(v_{n\alpha}({\bs r})\right)$  is the electron (hole) component of the eigenstates of  the single particle BdG Hamiltonian $\tilde{H}_{BdG}$ associated with the above many-body Hamiltonian in the rotating frame [see Eq. 2 in the Main Text]. The spin operator can be expressed  in terms of the bogoliubons and the single-particle wavefunctions:
\begin{align}
\hat{\bs \sigma}({\bs 0})&=\sum_{n,\alpha,\beta}\left[u^*_{n\alpha}({\bs 0})\gamma_n^\dagger-\alpha v_{n\alpha}({\bs 0})\gamma_n\right]({\bs\sigma})_{\alpha\beta}\left[u_{n\beta}({\bs 0})\gamma_n-\beta v^*_{n\beta}({\bs 0})\gamma^\dagger_n\right]\nn\\
&=\sum_{n,\alpha,\beta}\left[\left(u^*_{n\alpha}({\bs 0})({\bs\sigma})_{\alpha\beta}u_{n\beta}({\bs 0})-\alpha\beta v_{n\alpha}({\bs 0})({\bs\sigma})_{\alpha\beta}v^*_{n\beta}({\bs 0})\right)\gamma_n^\dagger\gamma_n+\alpha\beta v_{n\alpha}({\bs 0})({\bs\sigma})_{\alpha\beta}v^*_{n\beta}({\bs 0})\right]\,.
\label{SpinOp}
\end{align}

The BCS Hamiltonian above can be written in a compact way as:
\begin{equation}
H_{BCS}(t)=\frac{1}{2}\int d{\bs r}\,\Psi^\dagger({\bs r})H_{BdG}(t)\,\Psi({\bs r})\,,
\end{equation}
with $\Psi({\bs r})=\{c_\uparrow({\bs r}),c_\downarrow({\bs r}),c^\dagger_\downarrow({\bs r}), -c^\dagger_\uparrow({\bs r})\}^T$, and
\begin{equation}
H_{BdG}(t)=\left(\frac{p^2}{2m}-\mu\right)\tau_z+\Delta\tau_x-\underbrace{J\,{\bs S}(t)\cdot{\bs\sigma}\,\delta({\bs r})}_{V_{ i}}\,,
\end{equation}
being the single-particle  BdG Hamiltonian, with ${\bs \tau}=(\tau_x, \tau_y, \tau_z)$ being Pauli matrices acting in the particle-hole (Nambu) space.  We can then find the rotating frame version of this Hamiltonian by performing  a  unitary transformation $U(t)=e^{-i\sigma_zbt}$, or
\begin{equation}
\tilde H_{BdG}=U^\dagger(t)H_{BdG}(t)U(t)+i\dot{U}^\dagger(t)U(t)\,,
\end{equation}
which in turn leads to
\begin{align}
H_{BdG}(t)&\rightarrow \tilde{H}_{BdG}=H_{BdG}(0)-b\sigma_z\,.
\end{align}
In order to evaluate the eigenvalues and eigenstates of the BdG Hamiltonian,  we write down the  bare Green's function at ${\bs r}=0$ in the rotating frame:
\begin{equation}
\tilde G_0(\omega)=-\frac{\pi\nu_0}{2}\sum_{\sigma=\pm1}\frac{\omega+\sigma b+\Delta\tau_x}{\sqrt{\Delta^2-(\omega+\sigma b)^2}}(1+\sigma\sigma_z)\,,
\end{equation}
from which the energy of the Shiba states can be calculated from the following eigenvalue equation:
\begin{equation}
[I+\tilde{G}_0(\omega,{\bs r}=0)\,J\,{\bs S}(0)\cdot{\bs \sigma}]\Phi_S({\bs 0})=0\,,
\end{equation}
with $\Phi_S({\bs 0})=\{u_{0\uparrow}, u_{0\downarrow}, v_{0\downarrow}, -v_{0\uparrow}\}^T$ being the wavefunction of the Shiba state with the corresponding particle ($u_{0\sigma}$) and hole ($v_{0\sigma}$) components.

In the limit $b/\Delta\ll1$ and $\alpha\approx 1$ (deep Shiba limit), we obtain
\begin{equation}
E_S'\approx E_S-b\cos\theta\,,
\label{ShibaE}
\end{equation}
while for the wavefunctions, in the same order, we find:
\begin{equation}
\Phi_S({\bf 0})\approx\frac{1}{\sqrt{N}}\left[\begin{pmatrix}
\cos(\theta/2)\\
\sin(\theta/2)\\
\cos(\theta/2)\\
\sin(\theta/2)
\end{pmatrix}+\frac{b}{\Delta}\frac{(1+\alpha^2)^2\sin\theta}{8\alpha^2}
\begin{pmatrix}
\sin(\theta/2)\\
-\cos(\theta/2)\\
\sin(\theta/2)\\
-\cos(\theta/2)
\end{pmatrix}
\right]\,.
\label{seq:5}
\end{equation} 
We mention that in order to fully account for all the leading terms in $b/\Delta$, we expanded both the wave function at the position of the impurity and the normalization factor, with $N=(1+\alpha^2)^2/(2\pi\nu_0\alpha\Delta)$ being  the bare wave function normalization factor (i.e. in the absence of precession).  From the wave function we see that the dynamics induces coupling between the static Shiba state to its spin partner in the continuum. The above result can be interpreted in the context of an adiabatically driven spin $1/2$ particle: when subjected to a time-dependent magnetic field ${\bs B}(t)$ (constant in magnitude, $|{\bs B}(t)|\equiv B$), the lowest instantaneous  eigenstate, $|\psi_\uparrow\rangle=\left[\cos{(\theta/2)},\sin{(\theta/2})\right]^T$, gets renormalized by the dynamics as follows \cite{KOLODRUBETZ20171}:
\begin{equation}
|\psi'_\uparrow\rangle=|\psi_\uparrow\rangle+\frac{b}{2B}\sin\theta|\psi_\downarrow\rangle\,,
\end{equation}
where the state $|\psi_\downarrow\rangle=\left[\sin{(\theta/2)},-\cos{(\theta/2)}\right]^T$ is the higher energy spinor and they both satisfy the instantaneous Schrodinger equation ${\bf B}(t)\cdot{\bs \sigma}|\psi_\sigma\rangle=\sigma B|\psi_\sigma\rangle$. Eq.~\eqref{seq:5} has precisely the same form if we identify the effective  magnetic field separating the two states as:
\begin{equation}
B_{\rm eff}=\Delta\frac{4\alpha^2}{(1+\alpha^2)^2}\,.
\end{equation}
Thus, we can utilize an effective spin description to account for the non-adiabatic effects pertaining to the dynamical Shiba state at ${\bs r}={\bs 0}$. For $\alpha=1$ we find $B_{\rm eff}=\Delta$, i.e. the superconducting gap coincides with the effective magnetic field.  

Considering here only the effect of the Shiba states, $\gamma_0$, we can write:
\begin{equation}
c_{\alpha}({\bf 0})\approx u_{0\alpha}({\bf 0})\gamma_0-\alpha v^*_{0\alpha}({\bf 0})\gamma^\dagger_0\,,
\label{seq:10}
\end{equation}
where for the precessing case, from  Eq.~\eqref{seq:5},
\begin{align}
u_{0\uparrow}({\bf 0})=&v_{0\downarrow}({\bf 0})\approx\frac{1}{\sqrt{N}}\left(\cos\frac{\theta}{2}+\frac{X\sin\theta}{2}\sin\frac{\theta}{2}\right)\,,\nn\\
u_{0\downarrow}({\bf 0})=&v_{0\uparrow}({\bf 0})\approx\frac{1}{\sqrt{N}}\left(\sin\frac{\theta}{2}-\frac{X\sin\theta}{2}\cos\frac{\theta}{2}\right)\,,
\label{seq:17}
\end{align} 
with $X=\frac{2b}{B_{\rm eff}}=\frac{b}{\Delta}\frac{(1+\alpha^2)^2}{4\alpha^2}$.
For the general precessing case, the expectation values of the spin components in leading order in $b/\Delta$ can be calculated by using Eq.~\eqref{seq:17} in Eq.~\eqref{SpinOp}. In the laboratory frame, the components of spin average value at ${\bs r}=0$ are
\begin{align}
\langle\sigma_x(t)\rangle&\approx 2\left(n_S-\frac{1}{2}\right)\frac{1-X\cos{\theta}}{N}\cos(\Omega t)\sin{\theta}\,,\nn\\
\langle\sigma_y(t)\rangle&\approx2\left(n_S-\frac{1}{2}\right)\frac{1-X\cos{\theta}}{N}\sin(\Omega t)\sin\theta\,,\nn\\
\langle\sigma_z(t)\rangle&\approx2\left(n_S-\frac{1}{2}\right)\left(\frac{1-X\cos\theta}{N}\cos\theta+\frac{X}{N}\right)\,,
\end{align}
where $n_S=\langle\gamma^\dagger_0\gamma_0\rangle$ is the occupation number for the Shiba state. For $\theta=0$,
$\langle{\bs \sigma}(t)\rangle=(2/N)(n_S-1/2) {\bs z}$, i.e. aligned with the classical spin direction, which is the same situation for $b=0$ and arbitrary $\theta$, which pertains to  ${\bs z}\rightarrow{\bs n}$ in the above expression.

\subsubsection{Feedback on the classical magnetic moment}

The feedback effect of the Shiba state on the dynamics of the magnetic impurity spin can be revealed by calculating the torque induced on the classical spin 
up to leading order in the driving frequency $\Omega=b/2$. The torque acting on the magnetic impurity spin reads:
\begin{align}
{\bs \tau}_R(t)&=J {\bs S}(t)\times \langle{\bs \sigma}\delta({\bs r})\rangle
=\left(n_S-\frac{1}{2}\right)\frac{b}{S}{\bs S}(t)\times{\bs z}\,.
\end{align}
As mentioned in the Main Text, this equation shows that the torque on the magnetic impurity due to the Shiba states depends only on the precession frequency of the spin 
and the occupation number.

The Landau–Lifshitz-Gilbert equation describing the dynamics of the driven classical spin can now be written as 
\begin{align}
\dot{\bs S}(t)&=-\gamma{\bs S}(t)\times\left({\bs B}(t)+\beta \dot{\bs S}(t)\right)+{\bs\tau}_R(t)\nonumber\\
&=-{\bs S}(t)\times\left[\gamma{\bs B}(t)-\left(n_S-\frac{1}{2}\right)\frac{b}{S}\,{\bs z}+\beta \dot{\bs S}(t)\right]\,,
\end{align}
where $\gamma B_0\equiv \Omega_0$ defines the resonance frequency in the absence of the Shiba states. Given $B_0$, in the presence of the Shiba states, this resonance condition becomes:
\begin{align}
\Omega_0'=\Omega_0-\left(n_S-\frac{1}{2}\right)\frac{1}{2S}\Omega_0'\,,
\end{align}
which in turn allows us to relate the bare and renormalized frequencies as follows:
\begin{equation}
\Omega_0'=\displaystyle{\frac{\Omega_0}{\displaystyle{1+\left(n_S-\frac{1}{2}\right)\frac{1}{2S}}}}\,.
\end{equation}
We see that the Shiba state, thus, affects the ferromagnetic resonance of the impurity via a shift independent of $J$. For a filled Shiba state, we can calculate the frequency shift to be $\Omega_0/(4S+1)$. Such a contribution pertains to the Berry phase effects associated with the precession, and have been discussed in other settings before. Note, however, that these results are applicable in the limit $S\gg1$ and $b\ll \Delta$ and does not account for the out-of-equilibrium processes, the bulk states,  or the dissipation effects in the wave-function formalism. Hence, we switch to Green's function approach to address all these effects.

\subsubsection{Geometrical interpretation}

The above results can be obtained from a purely geometrical approach, following Ref.~\onlinecite{KOLODRUBETZ20171}. Instead of switching to the rotating frame, the idea is to  perform a time-dependent transformation into the frame where the impurity points along the $z$ direction. Note that the force acting on the magnet by the electrons can be casted as ${\bs f}_S=\langle\partial_{\bs S}H_{BdG}[{\bs S}(t)]\rangle$, and the corresponding torque is ${\bs \tau}_S={\bs S}\times{\bs f}_S$. In the following we show briefly how to evaluate this term using adiabatic perturbation. The transformation $U(t)$ that diagonalizes the $V_i$ in the $H_{BdG}(t)$ Hamiltonian and performs the above task reads:
\begin{align}
U(t)\equiv U(\theta,\phi)=\left(
\begin{array}{cc}
\cos{(\theta/2)} & \sin{(\theta/2)} \\
e^{i\phi}\sin{(\theta/2)} &  -e^{i\phi} \cos{(\theta/2)}
\end{array}
\right)\,,
\end{align}
where $\phi$ is the polar angle  of the classical spin, which for circular precession is just $\phi\equiv\Omega t$. This  results in the following Hamiltonian in the moving frame:
\begin{align}
\tilde{H}_{BdG}(t)&=\bar{H}_{BdG}-\dot{\bs S}\cdot{\bs A}_S\,,\\
\bar{H}_{BdG}&=H_{SC}-JS\sigma_z\,\delta({\bs r})\,,
\end{align}
where ${\bs A}_{S}$ is the resulting gauge potential pertaining to the dynamics of the spin ${\bs S}$. Note that for the present case, we can in fact find an analytical the following well-known  analytical expression:
\begin{align}
\dot{\bs S}\cdot{\bs A}_S\equiv-\frac{1}{2}(1-\cos{\theta}\sigma_z-\sin{\theta}\sigma_x)\dot{\phi}-\frac{1}{2}\sigma_y\dot{\theta}\,. 
\end{align}

The density matrix evolution  in the Heisenberg picture with respect to $\bar{H}_{BdG}$ reads:
\begin{align}
i\frac{d\tilde{\rho}_H(t)}{dt}=\dot{\bs S}\cdot \left[{\bs A}_{H,S}, \tilde{\rho}_H(t)\right]\,,
\end{align}	
which can be solved as: 
\begin{align}
\tilde{\rho}_H(t)&\approx \rho_0+i\int_0^td\tau\,\dot{\bs S}(\tau)\cdot\left[{\bs A}_{H,S}(\tau),\rho_0(\tau)\right]\,,	
\end{align}
where we kept only the leading order terms in the dynamics, with  $\rho_0$ being the density matrix in the absence of the off-diagonal terms stemming from the gauge field (however, the diagonal terms can affect the spectrum and thus $\rho_0$, which is assumed in the following).   
This density matrix can now be used to evaluate the expectation value for the force, supplemented by the relations \cite{KOLODRUBETZ20171}:
\begin{align}
\langle n|{\bs A}_S|m\rangle=i\frac{\langle n|\partial_{\bs S} H_{BdG}(t)|m\rangle}{\epsilon_n-\epsilon_m}\,,
\end{align}
which then gives:
\begin{align}
{\bs f}_{S}(t)&={\rm Tr}[\tilde{\rho}_H(t) \partial_{\bs S} H_{H,BdG}(t)]\approx {\bs e}_\beta\,\dot{S}_\alpha(t) \sum_{n}\rho_0^{n} \langle n|\mathcal{F}_{\alpha\beta}({\bs S})|n\rangle\equiv\bar{\mathcal{F}}({\bs S})\,\dot{\bs S}\,,
\end{align}
with  
\begin{align}
\mathcal{F}_{\alpha\beta}({\bs S})=\partial_\alpha A_S^\beta-\partial_\beta A_S^\alpha\,,
\end{align}
being the  Berry curvature associated with the level $n$ (note that the gauge fields are projected onto this subspace), and in the derivatives is to be understood that $\alpha=S_\alpha$. Moreover, $\bar{\mathcal{F}}({\bs S})$ is the average  Berry curvature tensor over the occupation of the many-body states $n$. Note that the Berry curvature is a symplectic form on ${\mathcal S}^2$ (sphere) and we can transform it to the usual magnetic field $\mathcal{B}_\gamma=(1/2)\epsilon_{\gamma\alpha\beta}\mathcal{F}_{\alpha\beta}$, and for the simple case above $ \mathcal{B}\equiv(\mathcal{B}_x, \mathcal{B}_y, \mathcal{B}_z)=(1/2S^2){\bs n}$, as it is the case for a spin $1/2$ (the $1/S^2$ factor occurs since we perform the derivatives with respect to the full spin $S$).   

Note that the above gauge fields depend only on the spin degree of freedom and are independent of the position.  
In order to evaluate the Berry curvature, we just need the instantaneous states and more specifically, the Shiba states pertaining to the lowest occupied  and unoccupied levels, respectively: 
\begin{equation}
|\Phi_S^{\uparrow}({\bf 0})\rangle=\frac{1}{\sqrt{N}}\begin{pmatrix}
1\\
0\\
1\\
0
\end{pmatrix}
;\,\,\,
|\Phi_S^{\downarrow}({\bf 0})\rangle=\frac{1}{\sqrt{N}}\begin{pmatrix}
0\\
1\\
0\\
-1
\end{pmatrix}
\label{seq:5}
\end{equation} 
and the full position dependent $|\Phi_S^\sigma({\bf r})\rangle=[A_\sigma({\bs r})+B_\sigma({\bs r})\tau_z+C_\sigma({\bs r})\tau_x]|\Phi_S^\sigma({\bf 0})\rangle$ does not affect the spinors (the functions $A_\sigma$, $B_\sigma$, and $C_\sigma$ are spin independent), but only mix the electron and hole components. Finally, we need to account for the occupation (or density matrix) of these levels, which in the low temperature limit satisfy $\rho_{0}^\uparrow+\rho_{0}^\downarrow\approx1$ (but note, however, that they can still depend on the dynamics via the diagonal contributions in velocity).  

Putting everything together, we finally obtain:
\begin{equation}
{\bs f}_S={\bs e}_\beta(\rho_0^\uparrow-\rho_0^\downarrow)\mathcal{F}^S_{\alpha\beta}\dot{S}_\alpha=(\rho_0^\uparrow-\rho_0^\downarrow)\epsilon_{\gamma\alpha\beta}\mathcal{B}_\gamma\dot{S}_\alpha{\bs e}_\beta\equiv \frac{1}{2S}\left(n_S-\frac{1}{2}\right)({\bs n}\times\dot{\bs n})\,,
\end{equation}
where we defined the Shiba occupation number as $n_S\equiv\rho_0^\uparrow$, and we can easily identify $F_s[{\bs n}]\equiv S\,  \mathcal{B}\cdot{\bs n}=1/2$ in the main text as the radial Berry curvature. Finally, we can insert this expression into the equation of motion for the magnet:
\begin{align}
\dot{{\bs n}}=-\gamma {\bs n}\times{\bs B}(t)+\frac{1}{2S}\left(n_S-\frac{1}{2}\right){\bs n}\times({\bs n}\times\dot{\bs n})=-\gamma {\bs n}\times{\bs B}(t)+\frac{1}{2S}\left(n_S-\frac{1}{2}\right)\dot{\bs n}\,,
\end{align}
the last term now being the reactive torque on the magnet from the Shiba state, and results into the same frequency shift as in the previous subsection for circular preccesion. We mention here in passing that in a more general setting when, for example, spin orbit interaction is present, as well as various types of pairings, the LL equation becomes:
\begin{align}
\dot{{\bs n}}&=-\gamma {\bs n}\times{\bs B}(t)+\frac{F_s[{\bs n}]}{2S}\left(n_S-\frac{1}{2}\right)\,\dot{\bs n}\,,
\end{align}
with $F_s[{\bs n}]$ dependent on ${\bs n}$ itself.

\subsection{Green's function Approach}

In this section, we account for the finite linewidth of the Shiba state, which implies we cannot use the wave-function description anymore. We employ  the Keldysh approach for out-of-equilibrium processes. The Dyson equation for the retarded/advanced and lesser GF respectively reads \cite{jauho1998}:
\begin{align}
G^{R(A)}(t,t')&=g^{R(A)}_0(t,t')+g^{R(A)}_0\odot(V_{i}+\Sigma^{R(A)})\odot G^{R(A)}(t,t')\,,\\
G^<(t,t')&=[1+G^R\odot(V_{i}+\Sigma^{R})]\odot g_0^<\odot[1+(V_{ i}+\Sigma^{A})\odot G^A](t,t')+G^R\odot\Sigma^<\odot G^A(t,t')\,,
\end{align}
where $\Sigma^{R/A/<}$ are the retarded/advanced/lesser self-energy associated with relaxation processes in the SC, the free lesser Keldysh GF $g_0^<(t,t')$ is that of an unperturbed superconductor (it will be defined  below), and all matrices are in the spin $\otimes$ Nambu space ($4\times4$). Moreover, the $\odot$ pertains to the {\it non-commutative}  time-convolution:
\begin{align}
[f\odot g](t,t')=\int dt''f(t,t'')g(t'',t')\,,
\end{align}
with the unit element $\hat{1}=\delta(t-t')$. The bare GF thus reads:
\begin{align}
g_0^<(t,t')=-n_F\odot(g_0^R-g_0^A)(t,t')\,,
\end{align}
where $n_F$ is the equilibrium distribution, while $g_0^{R(A)}$ are the bare retarded (advanced) GFs of the substrate at ${\bs r}=0$.  
While this equation is written in the laboratory frame, the idea is that we can translate every GF into the rotating frame like:
\begin{align}
\tilde{G}(t,t')=U^{\dagger}(t) G(t,t')U(t')\,,
\end{align}
and similarly for the distribution function we get \cite{TauberPRB10}:
\begin{equation}
\tilde{n}_F(t)=n_F(t)e^{-ib\sigma_zt}\,,
\end{equation}
which is assumed in equilibrium (no biases applied). 
Then, we get the following expression for the lesser GF in terms of the equilibrium distribution function and the retarded and advanced GFs:
\begin{align}
\tilde{G}^<(t,t')&=\tilde{g}_0^<(t,t')+\tilde{g}_0^<\odot V_{ i}\odot\tilde{G}^A(t,t')+\tilde{G}^R\odot V_{ i}\odot\hat{g}_0^<(t,t')+\tilde{G}^R\odot V_{ i}\odot\hat{g}_0^<\odot V_{ i}\odot\tilde{G}^A(t,t')\nn\\
&=-(\tilde{G}^R\odot\tilde{n}_F-\tilde{n}_F\odot\tilde{G}^A)(t,t')-\tilde{G}^R\odot(\tilde{n}_F\odot V_{ i}-V_{ i}\odot\tilde{n}_F)\odot\tilde{G}^A(t,t')\,,
\end{align}
or taking into account the self-energy $\Sigma$:
\begin{align}
\tilde{G}^<(t,t')&=-(\tilde{G}^R\odot\tilde{n}_F-\tilde{n}_F\odot\tilde{G}^A)(t,t')-\tilde{G}^R\odot\Sigma^<\odot\tilde{G}^A(t,t')\nonumber\\
&-\tilde{G}^R\odot[\tilde{n}_F\odot(V_{ i}+\Sigma^A)-(V_{ i}+\Sigma^R)\odot\tilde{n}_F]\odot\tilde{G}^A(t,t')\,.
\end{align}
For simplicity, we assume the processes that cause the self-energy (e.g. phonons) to be spin-independent, and consider the Dynes type of self-energy: 
\begin{equation}
\Sigma_{R,A}=\pm i\Gamma\,,
\end{equation}
with $\Gamma$ being a phenomenological broadening (we neglect here any real parts of the self-energy,  for simplicity). The dissipation term $\Gamma$ can have both intrinsic and extrinsic origin. For the intrinsic case, the relaxation process of the Shiba state can be through release of the excited quasiparticle into the continuum bulk states which depends on the temperature giving finite linewidth to the Shiba state.  In the rotating frame the full Dyson equation becomes 
\begin{align}
[\tilde{G}^{R,A}(\omega)]^{-1}=[G_{0}^{R,A}(\omega)]^{-1}-V_{i}\pm i\Gamma\,,
\end{align}
which in general can be solved as:
\begin{align}
\tilde{G}^{R,A}(\omega)=\frac{\pi\nu_0}{D^{R,A}(\omega,\alpha,b,\theta)}M^{R,A}(\omega,\alpha,b,\theta)\,,
\end{align}
where $D^{R,A}(\omega,\alpha,b,\theta)$ is a function that contains all the possible zeros (poles or branch cuts) associated with the combined system, while $M^{R,A}(\omega,\alpha,b,\theta)$ is a $4\times4$ matrix that depends only weakly on $\omega$. 
If we focus on frequencies around one of the Shiba poles, and assuming as well  the limit of small precession frequency, $b\ll\Delta$ we can write:
\begin{align}
\tilde{G}_{\pm}^{R,A}(\omega)&\approx\frac{\pi\nu_0}{\omega\mp E_S'\pm i\Gamma_S}M^{R,A}(\mp E_S',\alpha,b,\theta)\,,
\end{align}
where $E_S'$ is the effective Shiba energy in the presence of precession, and $\Gamma_S\propto \Gamma$ is the effective Shiba linewidth. 

Any observable at ${\bs r}=0$ can be evaluated from the lesser GF, which in the rotating frame (and in the frequency space) reads:
\begin{align}
\tilde{G}^<(\omega)&=-(\tilde{G}^R \tilde{n}_F-\tilde{n}_F\tilde{G}^A)-\tilde{G}^R(\tilde{n}_FV_{ i}-V_{i}\tilde{n}_F)\tilde{G}^A\,,
\end{align}
with $\tilde{n}_F(\omega,b)=f_{0}(\omega,b)+\sigma_zf_s(\omega,b)$ being a matrix of Fermi-Dirac distribution function.  Basically, in the rotating frame  the bare distribution functions for spin $\sigma$ are shifted by $\sigma b$. In these calculations we do not solve for the distribution function self-consistently since we assume the adiabatic regime with frequencies $b\ll\Delta$. The instantaneous spin expectation value in the rotating frame can be written as:
\begin{equation}
\langle\tilde{\bs\sigma}(\bs 0)\rangle=-i\int \frac{d\omega}{2\pi} {\rm Tr}\left[\left({\bs \sigma}\otimes\frac{1+\tau_z}{2}\right)\tilde{G}^<(\omega)\right]\,,
\end{equation}
with  the term $(1+\tau_z)/2$ being introduced in order to account for the electron components only since any physical observable is related to  either the electrons or holes. The evaluation of the spin for the entire range of parameters can be done only numerically. However, some insight can be gained exploring the deepn Shiba limit, and in particular the contribution pertaining to the resulting Shiba state. 

For completeness, below we provide the full retarded GF in the absence of dissipation, as we can find a compact analytical expression for both  $D(\omega,\alpha,b,\theta)$ and $M^{R,A}(\omega,\alpha,b,\theta)$. The former reads:
\begin{align}
\label{geq15}
D(\omega,\alpha,b,\theta)=&\omega_1\omega_2(1+\alpha^4-2\alpha^2\cos^2\theta)-2\alpha\cos\theta(1-\alpha^2)[(b+\omega)\omega_2+(b-\omega)\omega_1]\nn\\
&+2\alpha^2[(b^2-\omega^2)(1+\cos^2\theta)-\Delta^2\sin^2\theta]\,,
\end{align}
where $\omega_1=\sqrt{\Delta^2-(\omega+b)^2}$ and $\omega_2=\sqrt{\Delta^2-(\omega-b)^2}$.
Solving $D(\omega,\alpha,b,\theta)=0$, we get the poles of the retarded GF which coincide with the renormalized  Shiba energies. 
The components of the $4\times4$ matrix $M(\omega,\alpha,b,\theta)$  read:
\begin{align}
M_{11}&=M_{33}=\omega_2(b+\omega)(\alpha^2\cos^2\theta-1)+\alpha^2\omega_1(b-\omega)(1+\cos^2\theta)+2\alpha(b^2-\omega^2)\cos\theta\,-\omega_1\omega_2\alpha(1-\alpha^2)\cos\theta\,,\nn\\
M_{22}&=M_{44}=\omega_1(\omega-b)(\alpha^2\cos^2\theta-1)-\alpha^2\omega_2(b+\omega)(1+\cos^2\theta)-2\alpha(b^2-\omega^2)\cos\theta\,
+\omega_1\omega_2\alpha(1-\alpha^2)\cos\theta\,,\nn\\
M_{12}&=M_{21}=M_{34}=M_{43}=\alpha\sin\theta\Big[b^2-\Delta^2-\omega^2+\alpha^2\omega_1\omega_2+\alpha\cos\theta\left(\omega_1(b-\omega)+\omega_2(b+\omega)\right)\Big]\,,\nn\\
M_{13}&=M_{31}=-\Delta\Big[\omega_2(1-\alpha^2\cos^2\theta)-\omega_1\alpha^2\sin^2\theta-2\alpha\cos\theta(b-\omega)\Big]\,,\nn\\
M_{14}&=M_{41}=M_{23}=M_{32}=-\Delta\Big[2\alpha\omega\sin\theta+\frac{\alpha^2}{2}\sin(2\theta)(\omega_1-\omega_2)\Big]\,,\nn\\
M_{24}&=M_{42}=-\Delta\Big[\omega_1(1-\alpha^2\cos^2\theta)-\omega_2\alpha^2\sin^2\theta-2\alpha\cos\theta(b+\omega)\Big]\,.
\end{align}
The last three lines pertain to the elements that represent the anomalous part of $\tilde{G}^R(\omega)$. As mentioned in the main text, the  anomalous GF can be written  as
\begin{equation}
\tilde{G}^{eh}(\omega)=A(\omega)\sigma_0+B(\omega)\sigma_z+C(\omega)\sigma_x\,,
\label{eq:15}
\end{equation}
where 
\begin{align}
A(\omega)=&-\frac{\pi\nu_0\Delta}{D(\omega)}\left[\frac{(\omega_1+\omega_2)(1-\alpha^2)-4b\alpha\cos\theta}{2}\right]\,,\nn\\
B(\omega)=&-\frac{\pi\nu_0\Delta}{D(\omega)}\left[\frac{(\omega_2-\omega_1)(1-\alpha^2\cos(2\theta))+4\omega\alpha\cos\theta}{2}\right]\,,\nn\\
C(\omega)=&-\frac{\pi\nu_0\Delta}{D(\omega)}\left[4\omega\sin\theta+\alpha(\omega_1-\omega_2)\sin(2\theta)\right]\,.
\end{align}
While $D(\omega)$ is even under $\omega\rightarrow -\omega$, we can separate the anomalous Green's function matrix elements into odd frequency pairing terms, $B(\omega),~C(\omega)$, and even frequency pairing terms, $A(\omega)$.

\subsubsection{Average spin value in deep Shiba limit and $b/\Delta\ll1$}

Considering the deep Shiba limit $\alpha\rightarrow1$ and the adiabatic condition $b/\Delta\ll1$, we can focus near the Shiba poles by  considering $\omega\approx \pm E'_S$. In leading order in $b$, the expectations values for the spin due to the in-gap Shiba states and   along the classical spin direction  reads (and which coincides with the average Shiba occupation number):
\begin{align}
\langle\sigma_{a,S}\rangle&\approx\frac{2}{\pi N}\left[\arctan\left(\frac{E_S+2b\sin^2{(\theta/2)}}{\Gamma_S}\right)\cos^2{(\theta/2)}+\arctan\left(\frac{E_S-2b\cos^2{(\theta/2)}}{\Gamma_S}\right)\sin^2{(\theta/2)}\right]\nonumber\\
&\equiv -\frac{2}{N}(n_S-1/2)\,,
\end{align}
while the perpendicular components, pertaining to the reactive and dissipative torques, respectively,  become:
\begin{align}
\langle\sigma_{a,R}\rangle&\approx\frac{b}{\pi}\left[\arctan\left(\frac{E_S+2b\sin^2{(\theta/2)}}{\Gamma_S}\right)\cos^2{(\theta/2)}+\arctan\left(\frac{E_S-2b\cos^2{(\theta/2)}}{\Gamma_S}\right)\sin^2{(\theta/2)}\right]\sin\theta\nonumber\\
&\equiv -b\,(n_S-1/2)\sin{\theta}\,,\\
\langle\sigma_{a,D}\rangle&\approx\frac{\Gamma_S}{2\pi}\left[\arctan\left(\frac{E_S+2b\sin^2{(\theta/2)}}{\Gamma_S}\right)-\arctan\left(\frac{E_S-2b\cos^2{(\theta/2)}}{\Gamma_S}\right)\right]\sin\theta\,,
\end{align}
where 
\begin{equation}
\Gamma_S=\frac{4\pi\alpha}{(1+\alpha^2)^2}\Gamma\equiv \frac{2}{N}\,\Gamma,\,
\end{equation}
is the effective Shiba linewidth. Comparison between the full numerics and the approximate solutions show good qualitative agreement in the limit of small $b$. Above, we utilized the full expression for the lesser GF, and expanded the resulting expectation values. Comparing with the wave function approach, we define
\begin{equation}
n_S=\frac{1}{2\pi}\left[\pi-\arctan\left(\frac{E_S+2b\sin^2{(\theta/2)}}{\Gamma_S}\right)\cos^2{(\theta/2)}-\arctan\left(\frac{E_S-2b\cos^2{(\theta/2)}}{\Gamma_S}\right)\sin^2{(\theta/2)}\right]\,.
\end{equation}

We mention that $\langle\sigma_S\rangle$ and $\langle\tau_R\rangle$  above can be obtained from the approximate Shiba GF (defining the spin quantization axis along the instantaneous classical spin direction):
\begin{align}
\tilde{G}^R_\pm(\omega)\approx&\frac{\pi\nu_0(\tau_0\pm\tau_x)}{\omega\mp E_S'-i\Gamma_S}\left[\frac{\alpha}{(1+\alpha^2)^2}\sigma_0\pm\left(\frac{\alpha\cos{\theta}}{(1+\alpha^2)^2}+\frac{b\sin^2{\theta}}{4\alpha}\right)\sigma_z\pm\left(\frac{\alpha\sin{\theta}}{(1+\alpha^2)^2}-\frac{b\sin{2\theta}}{8\alpha}\right)\sigma_x\right]\,,
\end{align}
where $\pm$ stands for the  Shiba pole at positive/negative energy. Note, however, that using this approximate GF leads to a zero dissipative torque, which can only be extracted going beyond the usual Shiba effective GF model (for example, the effective description in Ref.~\onlinecite{ruby2015} does not suffice). 

\subsection{Plots of dynamical torques vs. $b$}
\begin{figure}
	\includegraphics[width=0.505\linewidth,trim=400mm 0mm 100mm 0mm,clip]{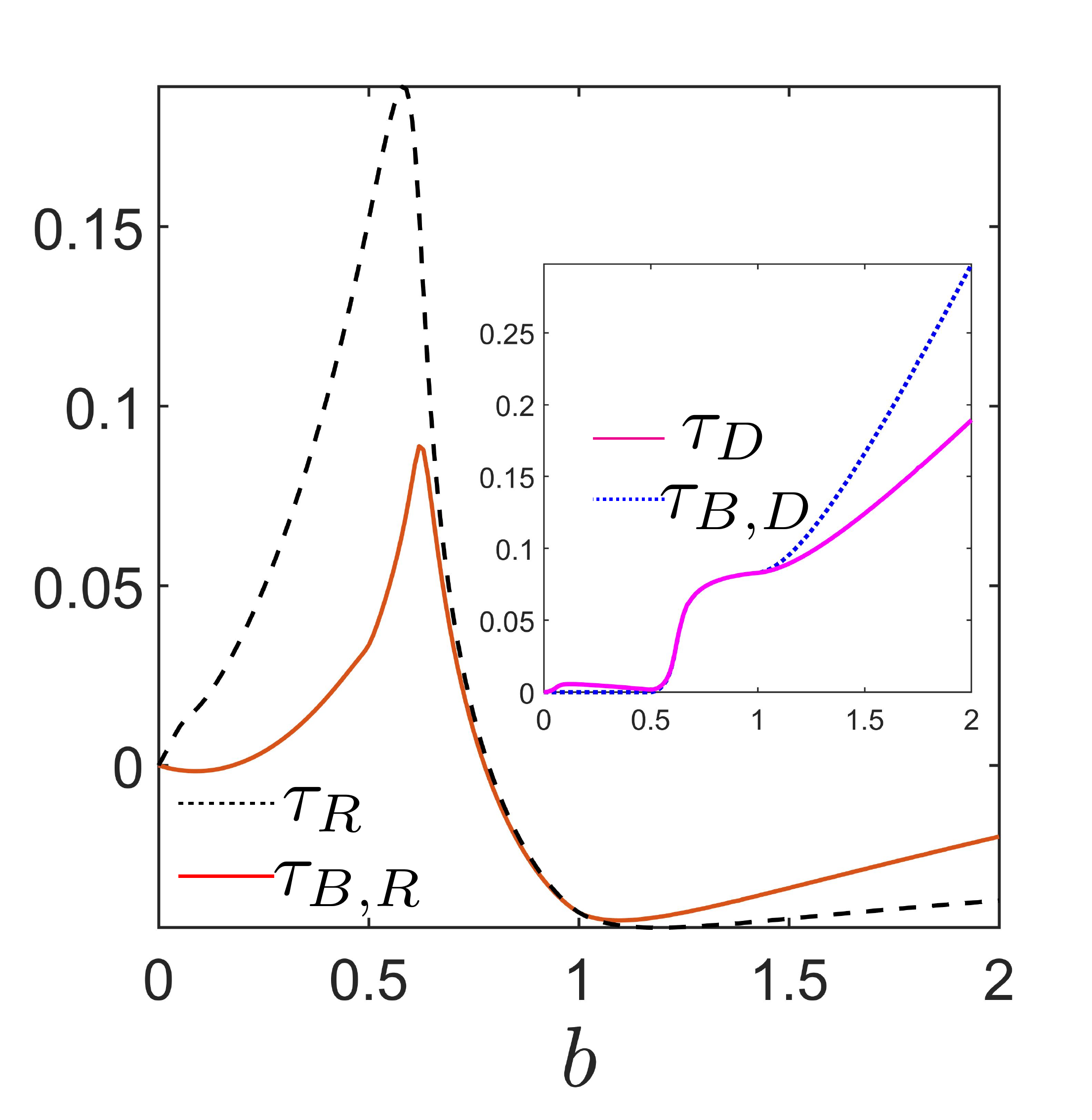}~~~~
	\includegraphics[width=0.48\linewidth,trim=400mm 0mm 100mm 0mm,clip]{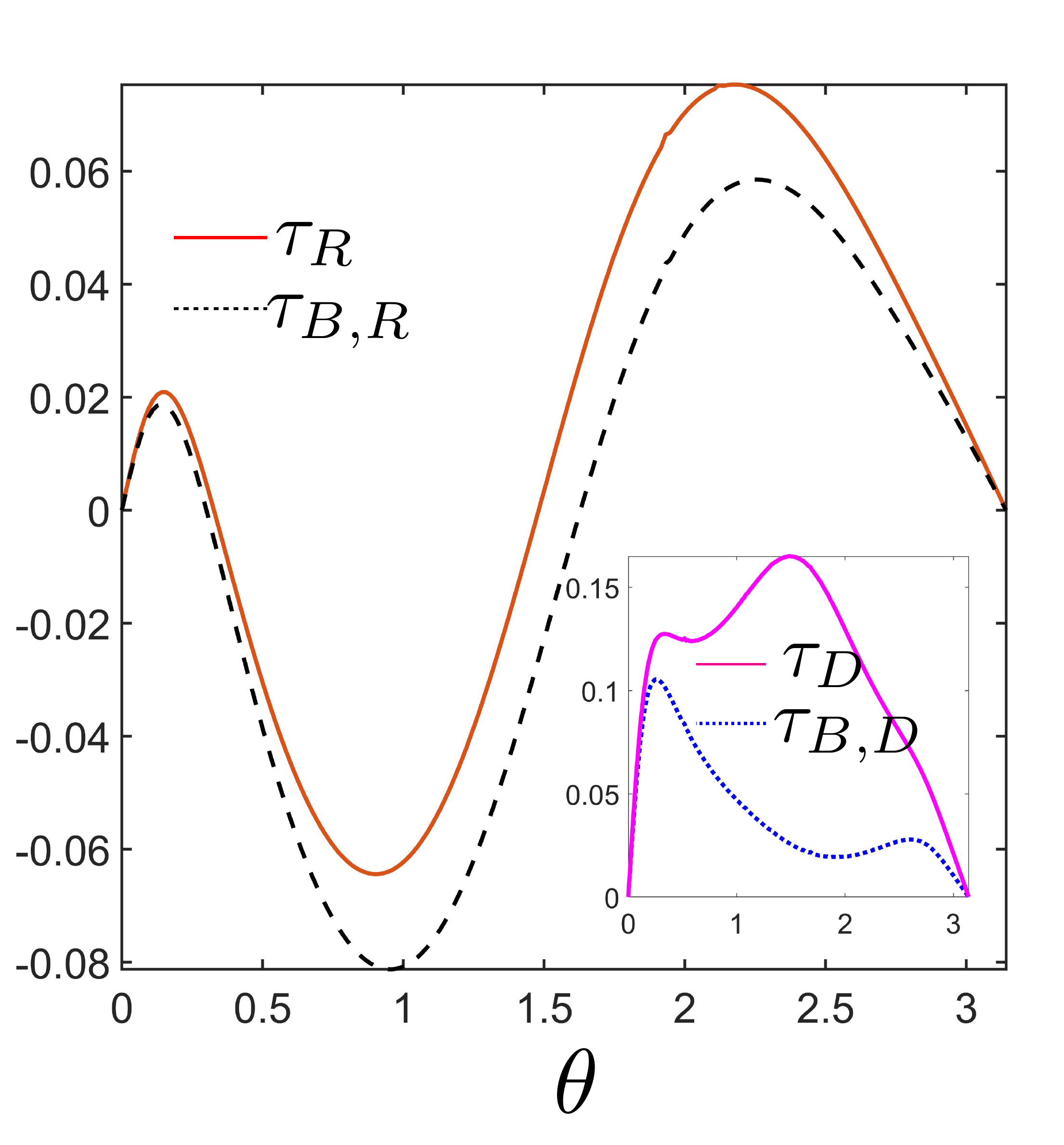}
	\caption{Variation of the reactive and dissipative torques. Main: Total $\tau_{R}$ (black dashed line) and the bulk $\tau_{B,R}$ (solid red) reactive torque respectively  (a)  as a function of  $b$,  for $\alpha=0.9$,  $\Gamma=0.01$, and  $\theta=\pi/6$ and (b)  as a function of $\theta$ for  $\alpha=0.9$,  $\Gamma=0.01$, and  $b=1.5$. The insets in the plots show the dependence of the full $\tau_{D}$ (solid magenta) and bulk $\tau_{B,D}$ (dotted blue) dissipative torques, respectively on the corresponding parameters, with the other values being the same as for the main plots.}
	\label{Fig:S1}
\end{figure}

While the main text shows the reactive and dissipative torques when the Shiba states are in the superconducting gap, Fig.~\ref{Fig:S1} shows the dynamic torques as function of $b$ for range upto $b=2>\Delta$ where the Shiba states enter the continuum. We find that the behavior of the torque in this regime is highly non-universal, as showed in the  plot presented this reply, as well as in the supplementary material. As $b$ is increased, the dynamic torques peak once the Shiba energy $|E_s|>\Delta-b$ where the Shiba state enters the continuum. While in the deep Shiba limit, the  contribution from the Shiba states to the dynamic torques on the classical spin dominates, as $b$ is increased bulk contribution increases and later dominates with large $b$. 


\begin{thebibliography}{36}
	\expandafter\ifx\csname natexlab\endcsname\relax\def\natexlab#1{#1}\fi
	\expandafter\ifx\csname bibnamefont\endcsname\relax
	\def\bibnamefont#1{#1}\fi
	\expandafter\ifx\csname bibfnamefont\endcsname\relax
	\def\bibfnamefont#1{#1}\fi
	\expandafter\ifx\csname citenamefont\endcsname\relax
	\def\citenamefont#1{#1}\fi
	\expandafter\ifx\csname url\endcsname\relax
	\def\url#1{\texttt{#1}}\fi
	\expandafter\ifx\csname urlprefix\endcsname\relax\def\urlprefix{URL }\fi
	\providecommand{\bibinfo}[2]{#2}
	\providecommand{\eprint}[2][]{\url{#2}}
	
	\bibitem[{\citenamefont{Sarma et~al.}(2015)\citenamefont{Sarma, Freedman, and
			Nayak}}]{sarma2015}
	\bibinfo{author}{\bibfnamefont{S.~D.} \bibnamefont{Sarma}},
	\bibinfo{author}{\bibfnamefont{M.}~\bibnamefont{Freedman}}, \bibnamefont{and}
	\bibinfo{author}{\bibfnamefont{C.}~\bibnamefont{Nayak}},
	\bibinfo{journal}{npj Quantum Information} \textbf{\bibinfo{volume}{1}},
	\bibinfo{pages}{1} (\bibinfo{year}{2015}).
	
	\bibitem[{\citenamefont{Choy et~al.}(2011)\citenamefont{Choy, Edge, Akhmerov,
			and Beenakker}}]{choy2011}
	\bibinfo{author}{\bibfnamefont{T. P.} \bibnamefont{Choy}},
	\bibinfo{author}{\bibfnamefont{J. M.}~\bibnamefont{Edge}},
	\bibinfo{author}{\bibfnamefont{A. R.}~\bibnamefont{Akhmerov}}, \bibnamefont{and}
	\bibinfo{author}{\bibfnamefont{C. W. J.}~\bibnamefont{Beenakker}},
	\bibinfo{journal}{Physical Review B} \textbf{\bibinfo{volume}{84}},
	\bibinfo{pages}{195442} (\bibinfo{year}{2011}).
	
	\bibitem[{\citenamefont{Nakosai et~al.}(2013)\citenamefont{Nakosai, Tanaka, and
			Nagaosa}}]{nakosai2013}
	\bibinfo{author}{\bibfnamefont{S.}~\bibnamefont{Nakosai}},
	\bibinfo{author}{\bibfnamefont{Y.}~\bibnamefont{Tanaka}}, \bibnamefont{and}
	\bibinfo{author}{\bibfnamefont{N.}~\bibnamefont{Nagaosa}},
	\bibinfo{journal}{Physical Review B} \textbf{\bibinfo{volume}{88}},
	\bibinfo{pages}{180503(R)} (\bibinfo{year}{2013}).
	
	\bibitem[{\citenamefont{Nadj-Perge et~al.}(2013)\citenamefont{Nadj-Perge,
			Drozdov, Bernevig, and Yazdani}}]{nadj2013}
	\bibinfo{author}{\bibfnamefont{S.}~\bibnamefont{Nadj-Perge}},
	\bibinfo{author}{\bibfnamefont{I. K.}~\bibnamefont{Drozdov}},
	\bibinfo{author}{\bibfnamefont{B.~A.} \bibnamefont{Bernevig}},
	\bibnamefont{and} \bibinfo{author}{\bibfnamefont{A.}~\bibnamefont{Yazdani}},
	\bibinfo{journal}{Physical Review B} \textbf{\bibinfo{volume}{88}},
	\bibinfo{pages}{020407(R)} (\bibinfo{year}{2013}).
	
	\bibitem[{\citenamefont{Braunecker and Simon}(2013)}]{braunecker2013}
	\bibinfo{author}{\bibfnamefont{B.}~\bibnamefont{Braunecker}} \bibnamefont{and}
	\bibinfo{author}{\bibfnamefont{P.}~\bibnamefont{Simon}},
	\bibinfo{journal}{Physical review letters} \textbf{\bibinfo{volume}{111}},
	\bibinfo{pages}{147202} (\bibinfo{year}{2013}).
	
	\bibitem[{\citenamefont{Klinovaja et~al.}(2013)\citenamefont{Klinovaja, Stano,
			Yazdani, and Loss}}]{klinovaja2013}
	\bibinfo{author}{\bibfnamefont{J.}~\bibnamefont{Klinovaja}},
	\bibinfo{author}{\bibfnamefont{P.}~\bibnamefont{Stano}},
	\bibinfo{author}{\bibfnamefont{A.}~\bibnamefont{Yazdani}}, \bibnamefont{and}
	\bibinfo{author}{\bibfnamefont{D.}~\bibnamefont{Loss}},
	\bibinfo{journal}{Physical review letters} \textbf{\bibinfo{volume}{111}},
	\bibinfo{pages}{186805} (\bibinfo{year}{2013}).
	
	\bibitem[{\citenamefont{Vazifeh and Franz}(2013)}]{vazifeh2013}
	\bibinfo{author}{\bibfnamefont{M. M.}~\bibnamefont{Vazifeh}} \bibnamefont{and}
	\bibinfo{author}{\bibfnamefont{M.}~\bibnamefont{Franz}},
	\bibinfo{journal}{Physical review letters} \textbf{\bibinfo{volume}{111}},
	\bibinfo{pages}{206802} (\bibinfo{year}{2013}).
	
	\bibitem[{\citenamefont{Pientka et~al.}(2013)\citenamefont{Pientka, Glazman,
			and von Oppen}}]{pientka2013}
	\bibinfo{author}{\bibfnamefont{F.}~\bibnamefont{Pientka}},
	\bibinfo{author}{\bibfnamefont{L.~I.} \bibnamefont{Glazman}},
	\bibnamefont{and} \bibinfo{author}{\bibfnamefont{F.}~\bibnamefont{von
			Oppen}}, \bibinfo{journal}{Physical Review B} \textbf{\bibinfo{volume}{88}},
	\bibinfo{pages}{155420} (\bibinfo{year}{2013}).
	
	\bibitem[{\citenamefont{Pientka et~al.}(2014)\citenamefont{Pientka, Glazman,
			and von Oppen}}]{pientka2014}
	\bibinfo{author}{\bibfnamefont{F.}~\bibnamefont{Pientka}},
	\bibinfo{author}{\bibfnamefont{L.~I.} \bibnamefont{Glazman}},
	\bibnamefont{and} \bibinfo{author}{\bibfnamefont{F.}~\bibnamefont{von
			Oppen}}, \bibinfo{journal}{Physical Review B} \textbf{\bibinfo{volume}{89}},
	\bibinfo{pages}{180505(R)} (\bibinfo{year}{2014}).
	
	\bibitem[{\citenamefont{P{\"o}yh{\"o}nen
			et~al.}(2014)\citenamefont{P{\"o}yh{\"o}nen, Weststr{\"o}m, R{\"o}ntynen, and
			Ojanen}}]{poyhonen2014}
	\bibinfo{author}{\bibfnamefont{K.}~\bibnamefont{P{\"o}yh{\"o}nen}},
	\bibinfo{author}{\bibfnamefont{A.}~\bibnamefont{Weststr{\"o}m}},
	\bibinfo{author}{\bibfnamefont{J.}~\bibnamefont{R{\"o}ntynen}},
	\bibnamefont{and} \bibinfo{author}{\bibfnamefont{T.}~\bibnamefont{Ojanen}},
	\bibinfo{journal}{Physical Review B} \textbf{\bibinfo{volume}{89}},
	\bibinfo{pages}{115109} (\bibinfo{year}{2014}).
	
	\bibitem[{\citenamefont{Heimes et~al.}(2014)\citenamefont{Heimes, Kotetes, and
			Sch{\"o}n}}]{heimes2014}
	\bibinfo{author}{\bibfnamefont{A.}~\bibnamefont{Heimes}},
	\bibinfo{author}{\bibfnamefont{P.}~\bibnamefont{Kotetes}}, \bibnamefont{and}
	\bibinfo{author}{\bibfnamefont{G.}~\bibnamefont{Sch{\"o}n}},
	\bibinfo{journal}{Physical Review B} \textbf{\bibinfo{volume}{90}},
	\bibinfo{pages}{060507(R)} (\bibinfo{year}{2014}).
	
	\bibitem[{\citenamefont{Reis et~al.}(2014)\citenamefont{Reis, Marchand, and
			Franz}}]{reis2014}
	\bibinfo{author}{\bibfnamefont{I.}~\bibnamefont{Reis}},
	\bibinfo{author}{\bibfnamefont{D.~ J.~J.}~\bibnamefont{Marchand}}, \bibnamefont{and}
	\bibinfo{author}{\bibfnamefont{M.}~\bibnamefont{Franz}},
	\bibinfo{journal}{Physical Review B} \textbf{\bibinfo{volume}{90}},
	\bibinfo{pages}{085124} (\bibinfo{year}{2014}).
	
	\bibitem[{\citenamefont{Weststr{\"o}m et~al.}(2015)\citenamefont{Weststr{\"o}m,
			P{\"o}yh{\"o}nen, and Ojanen}}]{weststrom2015}
	\bibinfo{author}{\bibfnamefont{A.}~\bibnamefont{Weststr{\"o}m}},
	\bibinfo{author}{\bibfnamefont{K.}~\bibnamefont{P{\"o}yh{\"o}nen}},
	\bibnamefont{and} \bibinfo{author}{\bibfnamefont{T.}~\bibnamefont{Ojanen}},
	\bibinfo{journal}{Physical Review B} \textbf{\bibinfo{volume}{91}},
	\bibinfo{pages}{064502} (\bibinfo{year}{2015}).
	
	\bibitem[{\citenamefont{Peng et~al.}(2015)\citenamefont{Peng, Pientka, Glazman,
			and von Oppen}}]{peng2015}
	\bibinfo{author}{\bibfnamefont{Y.}~\bibnamefont{Peng}},
	\bibinfo{author}{\bibfnamefont{F.}~\bibnamefont{Pientka}},
	\bibinfo{author}{\bibfnamefont{L.~I.} \bibnamefont{Glazman}},
	\bibnamefont{and} \bibinfo{author}{\bibfnamefont{F.}~\bibnamefont{von
			Oppen}}, \bibinfo{journal}{Physical review letters}
	\textbf{\bibinfo{volume}{114}}, \bibinfo{pages}{106801}
	(\bibinfo{year}{2015}).
	
	\bibitem[{\citenamefont{R{\"o}ntynen and Ojanen}(2015)}]{rontynen2015}
	\bibinfo{author}{\bibfnamefont{J.}~\bibnamefont{R{\"o}ntynen}}
	\bibnamefont{and} \bibinfo{author}{\bibfnamefont{T.}~\bibnamefont{Ojanen}},
	\bibinfo{journal}{Physical review letters} \textbf{\bibinfo{volume}{114}},
	\bibinfo{pages}{236803} (\bibinfo{year}{2015}).
	
	\bibitem[{\citenamefont{Braunecker and Simon}(2015)}]{braunecker2015}
	\bibinfo{author}{\bibfnamefont{B.}~\bibnamefont{Braunecker}} \bibnamefont{and}
	\bibinfo{author}{\bibfnamefont{P.}~\bibnamefont{Simon}},
	\bibinfo{journal}{Physical Review B} \textbf{\bibinfo{volume}{92}},
	\bibinfo{pages}{241410(R)} (\bibinfo{year}{2015}).
	
	\bibitem[{\citenamefont{Zhang et~al.}(2016)\citenamefont{Zhang, Kim, Rossi, and
			Lutchyn}}]{zhang2016}
	\bibinfo{author}{\bibfnamefont{J.}~\bibnamefont{Zhang}},
	\bibinfo{author}{\bibfnamefont{Y.}~\bibnamefont{Kim}},
	\bibinfo{author}{\bibfnamefont{E.}~\bibnamefont{Rossi}}, \bibnamefont{and}
	\bibinfo{author}{\bibfnamefont{R.~M.} \bibnamefont{Lutchyn}},
	\bibinfo{journal}{Physical Review B} \textbf{\bibinfo{volume}{93}},
	\bibinfo{pages}{024507} (\bibinfo{year}{2016}).
	
	\bibitem[{\citenamefont{Hoffman et~al.}(2016)\citenamefont{Hoffman, Klinovaja,
			and Loss}}]{hoffman2016}
	\bibinfo{author}{\bibfnamefont{S.}~\bibnamefont{Hoffman}},
	\bibinfo{author}{\bibfnamefont{J.}~\bibnamefont{Klinovaja}},
	\bibnamefont{and} \bibinfo{author}{\bibfnamefont{D.}~\bibnamefont{Loss}},
	\bibinfo{journal}{Physical Review B} \textbf{\bibinfo{volume}{93}},
	\bibinfo{pages}{165418} (\bibinfo{year}{2016}).
	
	\bibitem[{\citenamefont{Neupert et~al.}(2016)\citenamefont{Neupert, Yazdani,
			and Bernevig}}]{neupert2016}
	\bibinfo{author}{\bibfnamefont{T.}~\bibnamefont{Neupert}},
	\bibinfo{author}{\bibfnamefont{A.}~\bibnamefont{Yazdani}}, \bibnamefont{and}
	\bibinfo{author}{\bibfnamefont{B.~A.} \bibnamefont{Bernevig}},
	\bibinfo{journal}{Physical Review B} \textbf{\bibinfo{volume}{93}},
	\bibinfo{pages}{094508} (\bibinfo{year}{2016}).
	
	\bibitem[{\citenamefont{Kimme and Hyart}(2016)}]{kimme2016}
	\bibinfo{author}{\bibfnamefont{L.}~\bibnamefont{Kimme}} \bibnamefont{and}
	\bibinfo{author}{\bibfnamefont{T.}~\bibnamefont{Hyart}},
	\bibinfo{journal}{Physical Review B} \textbf{\bibinfo{volume}{93}},
	\bibinfo{pages}{035134} (\bibinfo{year}{2016}).
	
	\bibitem[{\citenamefont{Kaladzhyan
			et~al.}(2017{\natexlab{a}})\citenamefont{Kaladzhyan, Simon, and
			Trif}}]{kaladzhyan2017}
	\bibinfo{author}{\bibfnamefont{V.}~\bibnamefont{Kaladzhyan}},
	\bibinfo{author}{\bibfnamefont{P.}~\bibnamefont{Simon}}, \bibnamefont{and}
	\bibinfo{author}{\bibfnamefont{M.}~\bibnamefont{Trif}},
	\bibinfo{journal}{Physical Review B} \textbf{\bibinfo{volume}{96}},
	\bibinfo{pages}{020507(R)} (\bibinfo{year}{2017}{\natexlab{a}}).
	
	\bibitem[{\citenamefont{Andolina and Simon}(2017)}]{andolina2017}
	\bibinfo{author}{\bibfnamefont{G.~M.} \bibnamefont{Andolina}} \bibnamefont{and}
	\bibinfo{author}{\bibfnamefont{P.}~\bibnamefont{Simon}},
	\bibinfo{journal}{Physical Review B} \textbf{\bibinfo{volume}{96}},
	\bibinfo{pages}{235411} (\bibinfo{year}{2017}).
	
	\bibitem[{\citenamefont{Shiba}(1968)}]{shiba1968}
	\bibinfo{author}{\bibfnamefont{H.}~\bibnamefont{Shiba}},
	\bibinfo{journal}{Progress of theoretical Physics}
	\textbf{\bibinfo{volume}{40}}, \bibinfo{pages}{435} (\bibinfo{year}{1968}).
	
	\bibitem[{\citenamefont{Yu}(1965)}]{yu1965}
	\bibinfo{author}{\bibfnamefont{L.}~\bibnamefont{Yu}}, \bibinfo{journal}{Acta
		Phys. Sin} \textbf{\bibinfo{volume}{21}}, \bibinfo{pages}{75}
	(\bibinfo{year}{1965}).
	
	\bibitem[{\citenamefont{Rusinov}(1969)}]{rusinov1969}
	\bibinfo{author}{\bibfnamefont{A.}~\bibnamefont{Rusinov}},
	\bibinfo{journal}{Soviet Journal of Experimental and Theoretical Physics
		Letters} \textbf{\bibinfo{volume}{9}}, \bibinfo{pages}{85}
	(\bibinfo{year}{1969}).
	
	\bibitem[{\citenamefont{Stiles and Miltat}(2006)}]{stiles2006}
	\bibinfo{author}{\bibfnamefont{M.~D.} \bibnamefont{Stiles}} \bibnamefont{and}
	\bibinfo{author}{\bibfnamefont{J.}~\bibnamefont{Miltat}}, in
	\emph{\bibinfo{booktitle}{Spin dynamics in confined magnetic structures III}}
	(\bibinfo{publisher}{Springer}, \bibinfo{year}{2006}), pp.
	\bibinfo{pages}{225--308}.
	
	\bibitem[{\citenamefont{Sankey et~al.}(2006)\citenamefont{Sankey, Braganca,
			Garcia, Krivorotov, Buhrman, and Ralph}}]{sankey2006}
	\bibinfo{author}{\bibfnamefont{J. C.}~\bibnamefont{Sankey}},
	\bibinfo{author}{\bibfnamefont{P. M.}~\bibnamefont{Braganca}},
	\bibinfo{author}{\bibfnamefont{A. G. F.}~\bibnamefont{Garcia}},
	\bibinfo{author}{\bibfnamefont{I. N.}~\bibnamefont{Krivorotov}},
	\bibinfo{author}{\bibfnamefont{R. A.}~\bibnamefont{Buhrman}}, \bibnamefont{and}
	\bibinfo{author}{\bibfnamefont{D. C.}~\bibnamefont{Ralph}},
	\bibinfo{journal}{Physical review letters} \textbf{\bibinfo{volume}{96}},
	\bibinfo{pages}{227601} (\bibinfo{year}{2006}).
	
	\bibitem[{\citenamefont{Kaladzhyan
			et~al.}(2017{\natexlab{b}})\citenamefont{Kaladzhyan, Hoffman, and
			Trif}}]{kaladzhyan2017a}
	\bibinfo{author}{\bibfnamefont{V.}~\bibnamefont{Kaladzhyan}},
	\bibinfo{author}{\bibfnamefont{S.}~\bibnamefont{Hoffman}}, \bibnamefont{and}
	\bibinfo{author}{\bibfnamefont{M.}~\bibnamefont{Trif}},
	\bibinfo{journal}{Physical Review B} \textbf{\bibinfo{volume}{95}},
	\bibinfo{pages}{195403} (\bibinfo{year}{2017}{\natexlab{b}}).
	
	\bibitem[{\citenamefont{Millis
			et~al.}(1995)\citenamefont{Millis, Littlewood and
			Shraiman}}]{millis1995}
	\bibinfo{author}{\bibfnamefont{A.~J.}~\bibnamefont{Millis}},
	\bibinfo{author}{\bibfnamefont{P.~B.}~\bibnamefont{Littlewood}}, \bibnamefont{and}
	\bibinfo{author}{\bibfnamefont{B.~I.}~\bibnamefont{Shraiman}},
	\bibinfo{journal}{Physical Review Letters} \textbf{\bibinfo{volume}{74}},
	\bibinfo{pages}{5144} (\bibinfo{year}{1995}).
	
	\bibitem[{\citenamefont{Stahl
			et~al.}(2017)\citenamefont{Stahl and
			Potthoff}}]{stahl2017}
	\bibinfo{author}{\bibfnamefont{C.}~\bibnamefont{Stahl}},
	 \bibnamefont{and}
	\bibinfo{author}{\bibfnamefont{M.}~\bibnamefont{Potthoff}},
	\bibinfo{journal}{Physical Review Letters} \textbf{\bibinfo{volume}{119}},
	\bibinfo{pages}{227203} (\bibinfo{year}{2017}).
	
	\bibitem[{\citenamefont{Elbracht
			et~al.}(2020)\citenamefont{Elbracht, Michel and
			Potthoff}}]{elbracht2020}
	\bibinfo{author}{\bibfnamefont{M.}~\bibnamefont{Elbracht}},
	\bibinfo{author}{\bibfnamefont{S.}~\bibnamefont{Michel}},
	 \bibnamefont{and}
	\bibinfo{author}{\bibfnamefont{M.}~\bibnamefont{Potthoff}},
	\bibinfo{journal}{Physical Review Letters} \textbf{\bibinfo{volume}{124}},
	\bibinfo{pages}{197202} (\bibinfo{year}{2020}).
	
	\bibitem[{\citenamefont{Suresh
			et~al.}(2020)\citenamefont{Suresh, Bajpai and
			Nikoli{\'c}}}]{suresh2020}
	\bibinfo{author}{\bibfnamefont{A.}~\bibnamefont{Suresh}},
	\bibinfo{author}{\bibfnamefont{U.}~\bibnamefont{Bajpai}},
	 \bibnamefont{and}
	\bibinfo{author}{\bibfnamefont{B.~K.}~\bibnamefont{Nikoli{\'c}}},
	\bibinfo{journal}{Physical Review B} \textbf{\bibinfo{volume}{101}},
	\bibinfo{pages}{214412} (\bibinfo{year}{2020}).
	
	\bibitem[{\citenamefont{Bajpai
			et~al.}(2020)\citenamefont{Bajpai and
			Nikoli{\'c}}}]{bajpai2020}
		\bibinfo{author}{\bibfnamefont{U.}~\bibnamefont{Bajpai}},
	 \bibnamefont{and}
	\bibinfo{author}{\bibfnamefont{B.~K.}~\bibnamefont{Nikoli{\'c}}},
	\bibinfo{journal}{Physical Review Letters} \textbf{\bibinfo{volume}{125}},
	\bibinfo{pages}{187202} (\bibinfo{year}{2020}).
	
	\bibitem[{\citenamefont{Mishra et~al.}(2020)\citenamefont{Mishra, Takei,
			Simon, and Trif}}]{SM}
	\bibinfo{author}{\bibfnamefont{A.}~\bibnamefont{Mishra}},
	\bibinfo{author}{\bibfnamefont{S.}~\bibnamefont{Takei}},
	\bibinfo{author}{\bibfnamefont{P.}~\bibnamefont{Simon}}, \bibnamefont{and}
	\bibinfo{author}{\bibfnamefont{M.}~\bibnamefont{Trif}}
	(\bibinfo{year}{2020}).
	
	\bibitem[{\citenamefont{Teber et~al.}(2010)\citenamefont{Teber, Holmqvist, and
			Fogelstr{\"o}m}}]{teber2010}
	\bibinfo{author}{\bibfnamefont{S.}~\bibnamefont{Teber}},
	\bibinfo{author}{\bibfnamefont{C.}~\bibnamefont{Holmqvist}},
	\bibnamefont{and}
	\bibinfo{author}{\bibfnamefont{M.}~\bibnamefont{Fogelstr{\"o}m}},
	\bibinfo{journal}{Physical Review B} \textbf{\bibinfo{volume}{81}},
	\bibinfo{pages}{174503} (\bibinfo{year}{2010}).
	
	\bibitem[{\citenamefont{Ruby et~al.}(2015)\citenamefont{Ruby, Pientka, Peng,
			von Oppen, Heinrich, and Franke}}]{ruby2015}
	\bibinfo{author}{\bibfnamefont{M.}~\bibnamefont{Ruby}},
	\bibinfo{author}{\bibfnamefont{F.}~\bibnamefont{Pientka}},
	\bibinfo{author}{\bibfnamefont{Y.}~\bibnamefont{Peng}},
	\bibinfo{author}{\bibfnamefont{F.}~\bibnamefont{von Oppen}},
	\bibinfo{author}{\bibfnamefont{B.~W.} \bibnamefont{Heinrich}},
	\bibnamefont{and} \bibinfo{author}{\bibfnamefont{K.~J.}
		\bibnamefont{Franke}}, \bibinfo{journal}{Physical review letters}
	\textbf{\bibinfo{volume}{115}}, \bibinfo{pages}{197204}
	(\bibinfo{year}{2015}).
	
	\bibitem[{\citenamefont{Kaladzhyan et~al.}(2016)\citenamefont{Kaladzhyan, Bena,
			and Simon}}]{kaladzhyan2016}
	\bibinfo{author}{\bibfnamefont{V.}~\bibnamefont{Kaladzhyan}},
	\bibinfo{author}{\bibfnamefont{C.}~\bibnamefont{Bena}}, \bibnamefont{and}
	\bibinfo{author}{\bibfnamefont{P.}~\bibnamefont{Simon}},
	\bibinfo{journal}{Journal of Physics: Condensed Matter}
	\textbf{\bibinfo{volume}{28}}, \bibinfo{pages}{485701}
	(\bibinfo{year}{2016}).
	
	\bibitem[{\citenamefont{Kuzmanovski et~al.}(2019)\citenamefont{Kuzmanovski,
			Souto, and Balatsky}}]{kuzmanovski2019}
	\bibinfo{author}{\bibfnamefont{D.}~\bibnamefont{Kuzmanovski}},
	\bibinfo{author}{\bibfnamefont{R.~S.} \bibnamefont{Souto}}, \bibnamefont{and}
	\bibinfo{author}{\bibfnamefont{A.~V.} \bibnamefont{Balatsky}},
	\bibinfo{journal}{Physical Rev. B}
	\textbf{\bibinfo{volume}{101}}, \bibinfo{pages}{094505}
	(\bibinfo{year}{2020}).
	
	\bibitem[{\citenamefont{Perrin et~al.}(2019)\citenamefont{Perrin, M{\'e}nard,
			Brun, Cren, Civelli, and Simon}}]{perrin2019}
	\bibinfo{author}{\bibfnamefont{V.}~\bibnamefont{Perrin}},
	\bibinfo{author}{\bibfnamefont{G.~C.} \bibnamefont{M{\'e}nard}},
	\bibinfo{author}{\bibfnamefont{C.}~\bibnamefont{Brun}},
	\bibinfo{author}{\bibfnamefont{T.}~\bibnamefont{Cren}},
	\bibinfo{author}{\bibfnamefont{M.}~\bibnamefont{Civelli}}, \bibnamefont{and}
	\bibinfo{author}{\bibfnamefont{P.}~\bibnamefont{Simon}},
	\bibinfo{journal}{arXiv preprint arXiv:1912.11241 (accepted in PRL)}  (\bibinfo{year}{2019}).
	
	\bibitem[{\citenamefont{Sinova et~al.}(2015)\citenamefont{Sinova,
			Valenzuela, Wunderlich, Back, and Jungwirth}}]{valenzuelaRMP15}
	\bibinfo{author}{\bibfnamefont{J.}~\bibnamefont{Sinova}},
	\bibinfo{author}{\bibfnamefont{S.~O.} \bibnamefont{Valenzuela}}, 
	\bibinfo{author}{\bibfnamefont{J.} \bibnamefont{Wunderlich}},
	\bibinfo{author}{\bibfnamefont{C.~H.} \bibnamefont{Back}},
	\bibnamefont{and}
	\bibinfo{author}{\bibfnamefont{T.} \bibnamefont{Jungwirth}},
	\bibinfo{journal}{Reviews of Modern Physics}
	\textbf{\bibinfo{volume}{87}}, \bibinfo{pages}{1213}
	(\bibinfo{year}{2015}).
	
	\bibitem[{\citenamefont{Hatter et~al.}(2017)\citenamefont{Hatter,
			Heinrich, Rolf, and Franke}}]{hatter2017}
	\bibinfo{author}{\bibfnamefont{N.}~\bibnamefont{Hatter}},
	\bibinfo{author}{\bibfnamefont{B.~W.} \bibnamefont{Heinrich}}, 
	\bibinfo{author}{\bibfnamefont{D.} \bibnamefont{Rolf}},
		\bibnamefont{and}
	\bibinfo{author}{\bibfnamefont{K.~J.} \bibnamefont{Franke}},
	\bibinfo{journal}{Nature communications}
	\textbf{\bibinfo{volume}{8}}, \bibinfo{pages}{1--7}
	(\bibinfo{year}{2017}).
	
	\bibitem[{\citenamefont{{\v{Z}}itko }(2018)\citenamefont{{\v{Z}}itko}}]{vzitko2018}
	\bibinfo{author}{\bibfnamefont{R.}~\bibnamefont{{\v{Z}}itko}},
	\bibinfo{journal}{PhyB}
	\textbf{\bibinfo{volume}{536}}, \bibinfo{pages}{230--234}
	(\bibinfo{year}{2018}).
	
	\bibitem[{\citenamefont{Willke et~al.}(2018)\citenamefont{Willke,
			Paul, Natterer, Yang, Bae, Choi, Fern{\'a}ndez-Rossier, Heinrich and Lutz}}]{willke2018}
	\bibinfo{author}{\bibfnamefont{P.}~\bibnamefont{Willke}},
	\bibinfo{author}{\bibfnamefont{W.} \bibnamefont{Paul}}, 
	\bibinfo{author}{\bibfnamefont{F.~D.} \bibnamefont{Natterer}},
	\bibinfo{author}{\bibfnamefont{K.} \bibnamefont{Yang}},
	\bibinfo{author}{\bibfnamefont{Y.} \bibnamefont{Bae}},
	\bibinfo{author}{\bibfnamefont{T.} \bibnamefont{Choi}},
	\bibinfo{author}{\bibfnamefont{J.} \bibnamefont{Fern{\'a}ndez-Rossier}},
	\bibinfo{author}{\bibfnamefont{A.~J.} \bibnamefont{Heinrich}},
	\bibnamefont{and}
	\bibinfo{author}{\bibfnamefont{C.~P.} \bibnamefont{Lutz}},
	\bibinfo{journal}{Science advances}
	\textbf{\bibinfo{volume}{4}}, \bibinfo{pages}{eaaq1543}
	(\bibinfo{year}{2018}).
	
	\bibitem[{\citenamefont{Natterer et~al.}(2017)\citenamefont{Natterer, Yang,
			Paul, Willke, Choi, Greber, Heinrich, and Lutz}}]{natterer2017}
	\bibinfo{author}{\bibfnamefont{F.~D.} \bibnamefont{Natterer}},
	\bibinfo{author}{\bibfnamefont{K.}~\bibnamefont{Yang}},
	\bibinfo{author}{\bibfnamefont{W.}~\bibnamefont{Paul}},
	\bibinfo{author}{\bibfnamefont{P.}~\bibnamefont{Willke}},
	\bibinfo{author}{\bibfnamefont{T.}~\bibnamefont{Choi}},
	\bibinfo{author}{\bibfnamefont{T.}~\bibnamefont{Greber}},
	\bibinfo{author}{\bibfnamefont{A.~J.} \bibnamefont{Heinrich}},
	\bibnamefont{and} \bibinfo{author}{\bibfnamefont{C.~P.} \bibnamefont{Lutz}},
	\bibinfo{journal}{Nature} \textbf{\bibinfo{volume}{543}},
	\bibinfo{pages}{226} (\bibinfo{year}{2017}).
	
	\bibitem[{\citenamefont{Yang et~al.}(2019)\citenamefont{Yang, Paul, Natterer,
			Lado, Bae, Willke, Choi, Ferr\'on, Fern\'andez-Rossier, Heinrich
			et~al.}}]{yang2019}
	\bibinfo{author}{\bibfnamefont{K.}~\bibnamefont{Yang}},
	\bibinfo{author}{\bibfnamefont{W.}~\bibnamefont{Paul}},
	\bibinfo{author}{\bibfnamefont{F.~D.} \bibnamefont{Natterer}},
	\bibinfo{author}{\bibfnamefont{J.~L.} \bibnamefont{Lado}},
	\bibinfo{author}{\bibfnamefont{Y.}~\bibnamefont{Bae}},
	\bibinfo{author}{\bibfnamefont{P.}~\bibnamefont{Willke}},
	\bibinfo{author}{\bibfnamefont{T.}~\bibnamefont{Choi}},
	\bibinfo{author}{\bibfnamefont{A.}~\bibnamefont{Ferr\'on}},
	\bibinfo{author}{\bibfnamefont{J.}~\bibnamefont{Fern\'andez-Rossier}},
	\bibinfo{author}{\bibfnamefont{A.~J.} \bibnamefont{Heinrich}},
	\bibnamefont{et~al.}, \bibinfo{journal}{Phys. Rev. Lett.}
	\textbf{\bibinfo{volume}{122}}, \bibinfo{pages}{227203}
	(\bibinfo{year}{2019}).
\end{thebibliography}

\begin{thebibliography}{5}
	\expandafter\ifx\csname natexlab\endcsname\relax\def\natexlab#1{#1}\fi
	\expandafter\ifx\csname bibnamefont\endcsname\relax
	\def\bibnamefont#1{#1}\fi
	\expandafter\ifx\csname bibfnamefont\endcsname\relax
	\def\bibfnamefont#1{#1}\fi
	\expandafter\ifx\csname citenamefont\endcsname\relax
	\def\citenamefont#1{#1}\fi
	\expandafter\ifx\csname url\endcsname\relax
	\def\url#1{\texttt{#1}}\fi
	\expandafter\ifx\csname urlprefix\endcsname\relax\def\urlprefix{URL }\fi
	\providecommand{\bibinfo}[2]{#2}
	\providecommand{\eprint}[2][]{\url{#2}}
	
	\bibitem[{\citenamefont{Teber et~al.}(2010)\citenamefont{Teber, Holmqvist, and
			Fogelstr\"om}}]{TauberPRB10}
	\bibinfo{author}{\bibfnamefont{S.}~\bibnamefont{Teber}},
	\bibinfo{author}{\bibfnamefont{C.}~\bibnamefont{Holmqvist}},
	\bibnamefont{and}
	\bibinfo{author}{\bibfnamefont{M.}~\bibnamefont{Fogelstr\"om}},
	\bibinfo{journal}{Phys. Rev. B} \textbf{\bibinfo{volume}{81}},
	\bibinfo{pages}{174503} (\bibinfo{year}{2010}),
	\urlprefix\url{https://link.aps.org/doi/10.1103/PhysRevB.81.174503}.
	
	\bibitem[{\citenamefont{Balatsky et~al.}(2006)\citenamefont{Balatsky, Vekhter,
			and Zhu}}]{BalatskyRMP06}
	\bibinfo{author}{\bibfnamefont{A.~V.} \bibnamefont{Balatsky}},
	\bibinfo{author}{\bibfnamefont{I.}~\bibnamefont{Vekhter}}, \bibnamefont{and}
	\bibinfo{author}{\bibfnamefont{J.-X.} \bibnamefont{Zhu}},
	\bibinfo{journal}{Rev. Mod. Phys.} \textbf{\bibinfo{volume}{78}},
	\bibinfo{pages}{373} (\bibinfo{year}{2006}),
	\urlprefix\url{https://link.aps.org/doi/10.1103/RevModPhys.78.373}.
	
	\bibitem[{\citenamefont{Kolodrubetz et~al.}(2017)\citenamefont{Kolodrubetz,
			Sels, Mehta, and Polkovnikov}}]{KOLODRUBETZ20171}
	\bibinfo{author}{\bibfnamefont{M.}~\bibnamefont{Kolodrubetz}},
	\bibinfo{author}{\bibfnamefont{D.}~\bibnamefont{Sels}},
	\bibinfo{author}{\bibfnamefont{P.}~\bibnamefont{Mehta}}, \bibnamefont{and}
	\bibinfo{author}{\bibfnamefont{A.}~\bibnamefont{Polkovnikov}},
	\bibinfo{journal}{Physics Reports} \textbf{\bibinfo{volume}{697}},
	\bibinfo{pages}{1 } (\bibinfo{year}{2017}), ISSN \bibinfo{issn}{0370-1573},
	\bibinfo{note}{geometry and non-adiabatic response in quantum and classical
		systems},
	\urlprefix\url{http://www.sciencedirect.com/science/article/pii/S0370157317301989}.
	
	\bibitem[{\citenamefont{H.~Haug and Jauho}(1998)}]{jauho1998}
	\bibinfo{author}{\bibfnamefont{H.}~\bibnamefont{H.~Haug}} \bibnamefont{and}
	\bibinfo{author}{\bibfnamefont{A.-P.} \bibnamefont{Jauho}}, in
	\emph{\bibinfo{booktitle}{Quantum Kinetics in Transport and Optics of
			Semiconductors}} (\bibinfo{publisher}{Springer-Verlag, Berlin},
	\bibinfo{year}{1998}).
	
	\bibitem[{\citenamefont{Ruby et~al.}(2015)\citenamefont{Ruby, Pientka, Peng,
			von Oppen, Heinrich, and Franke}}]{ruby2015}
	\bibinfo{author}{\bibfnamefont{M.}~\bibnamefont{Ruby}},
	\bibinfo{author}{\bibfnamefont{F.}~\bibnamefont{Pientka}},
	\bibinfo{author}{\bibfnamefont{Y.}~\bibnamefont{Peng}},
	\bibinfo{author}{\bibfnamefont{F.}~\bibnamefont{von Oppen}},
	\bibinfo{author}{\bibfnamefont{B.~W.} \bibnamefont{Heinrich}},
	\bibnamefont{and} \bibinfo{author}{\bibfnamefont{K.~J.}
		\bibnamefont{Franke}}, \bibinfo{journal}{Physical Review Letters}
	\textbf{\bibinfo{volume}{115}}, \bibinfo{pages}{197204}
	(\bibinfo{year}{2015}).
	
\end{thebibliography}
\end{document}